\documentclass[10pt,journal,cspaper,compsoc]{IEEEtran}
\usepackage[latin9]{inputenc}
\usepackage{array}
\usepackage{verbatim}
\usepackage{multirow}
\usepackage{amsmath}
\usepackage{amssymb}
\usepackage{graphicx}
\usepackage{cite}
\usepackage{epstopdf}
\usepackage[ruled,vlined]{algorithm2e}
\graphicspath{image/}
\usepackage{color}
\usepackage[caption=false,font=footnotesize]{subfig}
\usepackage{ragged2e}
\usepackage{balance}

\definecolor{comment-blue}{rgb}{0,0,1}
\definecolor{comment-red}{rgb}{1,0,0}
\definecolor{comment-black}{rgb}{0,0,0}
\newcommand{\revise}[1]{\textnormal{\color{comment-black}{#1}}\unskip}
\newcommand{\reviseAgain}[1]{\textnormal{\color{comment-black}{#1}}\unskip}

\ifCLASSOPTIONcompsoc
\else
\fi
\ifCLASSINFOpdf
\else
\fi
\begin{document}
%
\title{AMUSE: Empowering Users for Cost-Aware Offloading with Throughput-Delay Tradeoffs}
%
%
%
%

\author{Youngbin Im,~\IEEEmembership{Student Member,~IEEE,}
        Carlee Joe-Wong,~\IEEEmembership{Student Member,~IEEE,}\\
        Sangtae Ha,~\IEEEmembership{Senior Member,~IEEE,}
        Soumya Sen,~\IEEEmembership{Member,~IEEE,}\\
        Ted ``Taekyoung'' Kwon,~\IEEEmembership{Member,~IEEE,}
        and Mung Chiang,~\IEEEmembership{Fellow,~IEEE}
\IEEEcompsocitemizethanks{
\IEEEcompsocthanksitem Y. Im and S. Ha are with the University of Colorado, Boulder, Colorado.\protect\\
E-mails: \{youngbin.im,sangtae.ha\}@colorado.edu 
\IEEEcompsocthanksitem C. Joe-Wong and M. Chiang are with Princeton University, Princeton, New Jersey.\protect\\
E-mails: \{cjoe,chiangm\}@princeton.edu
\IEEEcompsocthanksitem S. Sen is with the Carlson School of Management, University of Minnesota, MN.\protect\\
E-mail: ssen@umn.edu
\IEEEcompsocthanksitem T. Kwon is with Seoul National University, Seoul, Korea\protect\\
(corresponding author to provide phone: +82 2 880 9105; fax: +82 2 872 2045; e-mail: tkkown@snu.ac.kr).} 
\thanks{An earlier version of this paper appeared in the IEEE INFOCOM 2013 mini-conference.}
}

%
%

\markboth{IEEE Transactions on Mobile Computing, vol. 15, no. 5, pp. 1062-1076, May 1 2016}%
{Shell \MakeLowercase{\textit{et al.}}: Bare Demo of IEEEtran.cls for Computer Society Journals}
%


\IEEEcompsoctitleabstractindextext{%
\begin{abstract}
To cope with recent exponential increases in demand for mobile data, wireless Internet service providers (ISPs) are increasingly changing their pricing plans and deploying WiFi hotspots to offload their mobile traffic. However, these ISP-centric approaches for traffic management do not always match the interests of mobile users. Users face a complex, multi-dimensional tradeoff between cost, throughput, and delay in making their offloading decisions: while they may save money and receive a higher throughput by waiting for WiFi access, they may not wait for WiFi if they are sensitive to delay. To navigate this tradeoff, we develop AMUSE (Adaptive bandwidth Management through USer-Empowerment), a functional prototype of a practical, cost-aware WiFi offloading system that takes into account a user's throughput-delay tradeoffs and cellular budget constraint. Based on predicted future usage and WiFi availability, AMUSE decides which applications to offload to what times of the day. Since nearly all traffic flows from mobile devices are TCP flows, we introduce a new receiver-side bandwidth allocation mechanism to practically enforce the assigned rate of each TCP application. Thus, AMUSE users can optimize their bandwidth rates according to their own cost-throughput-delay tradeoff without relying on support from different apps' content servers. Through a measurement study of 20 smartphone users' traffic usage traces, we observe that though users already offload a large amount of some application types, our framework can offload a significant additional portion of users' cellular traffic.
We implement AMUSE on Windows 7 tablets and evaluate its effectiveness with 3G and WiFi usage data obtained from a trial with 37 mobile users. Our results show that AMUSE improves user utility; when compared with AMUSE, other offloading algorithms yield 14\% and 27\% lower user utilities for light and heavy users, respectively. Intelligently managing users' competing interests for cost, throughput, and delay can therefore improve their offloading decisions.
\end{abstract}

}

\maketitle

\IEEEdisplaynotcompsoctitleabstractindextext

%
\IEEEpeerreviewmaketitle

%
%
%
%
%

\section{Introduction}

\label{sec:Introduction}



Recent unprecedented increases in demand for mobile data traffic have begun to stress many mobile operators' networks: Cisco, for instance, predicts that mobile data traffic will grow at 61\% annually from 2013 to 2018, reaching 15.9 exabytes per month by 2018 \cite{CiscoVNI2014}. To cope with this surge in data usage, which is driven by applications such as mobile video, cloud services, and online magazines, many ISPs (Internet service providers) have adopted tiered pricing plans with monthly data caps to discourage heavy usage \cite{sen2012incentivizing}. To further reduce network traffic, many ISPs have also introduced supplementary networks such as WiFi hotspots or femtocells to offload their cellular traffic \cite{nycwifi,chandrasekhar2008femtocell,sen2013survey}. Such supplementary offerings introduce new challenges for users as they decide which parts of their traffic can be offloaded at what times.

\subsection{Empowering User Decisions}

Many data plans, especially in the U.S., charge large overage fees when users exceed a monthly usage cap. While offloading to WiFi reduces cellular data usage, thus saving users money on their data spending, they must also take into account WiFi's intermittent availability and higher throughput performance.  At some times, e.g., while out shopping, a user does not have immediate WiFi access and must wait for WiFi connectivity. The user then faces a choice:
\begin{itemize}
\item
\emph{Don't wait for WiFi:} The user must consume cellular data, using up some of his data cap, and may experience lower throughput than WiFi. However, she need not wait for data access, which is important for urgent applications, e.g. email.
\item
\emph{Wait for WiFi:} The user can save money and experience higher throughput, but must decide \emph{how long} to wait. Different applications can wait for different periods of time, e.g., cloud backups might be more delay tolerant than photo uploads to Facebook. Given each app's willingness to wait for some period of time, users must anticipate whether WiFi will be available at that time and decide whether the potential savings in data offloading and potential increase in throughput are worth the wait.

Waiting for WiFi also introduces the risk that apps waiting for WiFi must share the limited 3G bandwidth, should WiFi ultimately not be available. Some apps, such as videos, will require a large amount of bandwidth; their quality can suffer significantly if they must share bandwidth with other apps, e.g., cloud backups.\footnote{While our systems apply to any form of cellular data, e.g., 3G or LTE networks, we frame our discussion in terms of 3G data. LTE speeds can exceed WiFi, which makes the users' tradeoffs more complicated and AMUSE even more useful.}
\end{itemize}

Most users will not manually balance these competing factors in making offloading decisions.  Thus, we propose a user-side, automated WiFi offloading system called AMUSE (Adaptive bandwidth Management through USer EMpowerment) that intelligently navigates these tradeoffs for the user.  AMUSE utilizes WiFi access and application usage predictions to decide how long application sessions should wait for WiFi and, in case WiFi is not available, to optimally allocate 3G bandwidth among different apps. Building such a system poses both algorithmic and implementation challenges--not only must the user's tradeoff between cost, throughput quality, and delay be quantified and balanced, but 
we require a way to automatically enforce AMUSE's waiting for WiFi and sharing of 3G bandwidth. In solving these problems, we make the following contributions:
\begin{enumerate}

\item We develop a system for cost-aware WiFi offloading that exploits a user's delay tolerances for different applications and makes offloading decisions satisfying her throughput-delay tradeoffs and 3G budget constraints.

\item To enforce AMUSE's bandwidth allocation decisions for each application, we implement a practical receiver-side rate control algorithm for TCP.\footnote{We assume that download traffic makes up most of users' usage, so that the receiver is synonymous with the user.}  The algorithm is fully contained on and driven by end-user devices, making it suitable for practical deployment as it requires no modification of the TCP server side.

\item In order to analyze current mobile offloading patterns and the potential to offload more traffic from different apps, we conduct a measurement study using application usage data collected from 20 Android smartphone users for one week.\footnote{Throughout this work, ``app usage data'' refers to the volume of data used by each application, not the time duration of application usage.} The results reveal several facts that show offloading practice and possibility of smartphone users. 

\item We surveyed 100 participants in the U.S. to evaluate users' tradeoff between the cost of 3G usage and their willingness to wait for WiFi access.  We incorporate the resulting cost-throughput-delay tradeoff estimates into our model, and evaluate AMUSE's performance using these results and 3G and WiFi usage data collected from a trial with 37 mobile users.

\end{enumerate}

AMUSE is the first WiFi offloading system to fully account for cost, delay, and throughput in offloading traffic from 3G to WiFi. Other works have considered using WiFi offloading to reduce cost within a basic delay constraint, e.g., by using predictions of WiFi connectivity to improve offloading \cite{breadcrumbs} or allocating more WiFi bandwidth to users who are expected to leave the WiFi coverage area in a short amount of time~\cite{santhapuribytestogo}. Mobility can also enhance prefetching data over WiFi \cite{siris2013performance}. Wiffler~\cite{Augmenting3G} considers a more sophisticated model of different applications' delay tolerances, but does not consider different apps' bandwidth needs or their need to share 3G bandwidth.

To fully incorporate cost, delay, and throughput, we build an end-to-end mobile offloading system. In the next section, we describe AMUSE's components and the challenges of developing this end-user system.

\subsection{Components of AMUSE}

Figure \ref{fig:overview} gives an overview of AMUSE's components and their interactions.  The system architecture comprises four main modules: the User Interface, Bandwidth Optimizer, TCP Rate Controller, and App-Level Session Tracker.  The latter two modules reside in the kernel and are accordingly shaded darker in the control flow diagram (Fig. \ref{fig:control}); these enforce the offloading decisions made by the User Interface and Bandwidth Optimizer, which reside in the user-space.  To illustrate the system's full set of interactions, the Bandwidth Optimizer is split into three components: two prediction modules for app usage and WiFi availability, and an algorithm that computes utility-maximizing offloading decisions.

\subsubsection{User Interface}

As suggested by its name, AMUSE's User Interface interacts directly with the user, displaying the offloading decisions made as well as the user's app-level usage history.  The user may also set her preferences on the user interface, e.g., the maximum budget for 3G usage and delay tolerances for different applications.

\subsubsection{Bandwidth Optimizer}

The Bandwidth Optimizer makes offloading decisions for the user, given the preferences set by the user on the User Interface.  It consists of the three medium-shaded components in Fig. \ref{fig:control}: app usage prediction, WiFi access prediction, and a utility maximization algorithm.

AMUSE uses an adaptive user mobility model to predict WiFi availability at future times (Fig. \ref{fig:control}). The app usage prediction component allows AMUSE to calculate the expected savings from offloading an application session and to allocate 3G bandwidth to all active apps at any given time, giving more bandwidth to the apps with higher bandwidth requirements. 

The user's offloading decisions at any given time must take into account \emph{future} offloading decisions--for instance, a myopic algorithm may delay all sessions in the morning to 12 noon, if the probability of WiFi access at that time is high.  However, should WiFi not be available then, all of the delayed sessions would have to share the limited 3G bandwidth, or else wait even longer for WiFi.  Thus, at the beginning of each day, AMUSE optimizes over the entire rest of the day, using estimates of WiFi access probabilities and the size of application sessions at different times (e.g., hours).  It then refines this initial solution over the day to reflect the observed usage and WiFi access patterns.

\subsubsection{TCP Rate Controller and Session Tracker}

\begin{figure}[t]
\centering
\subfloat[System implementation architecture.]{\includegraphics[width = 0.4\textwidth]{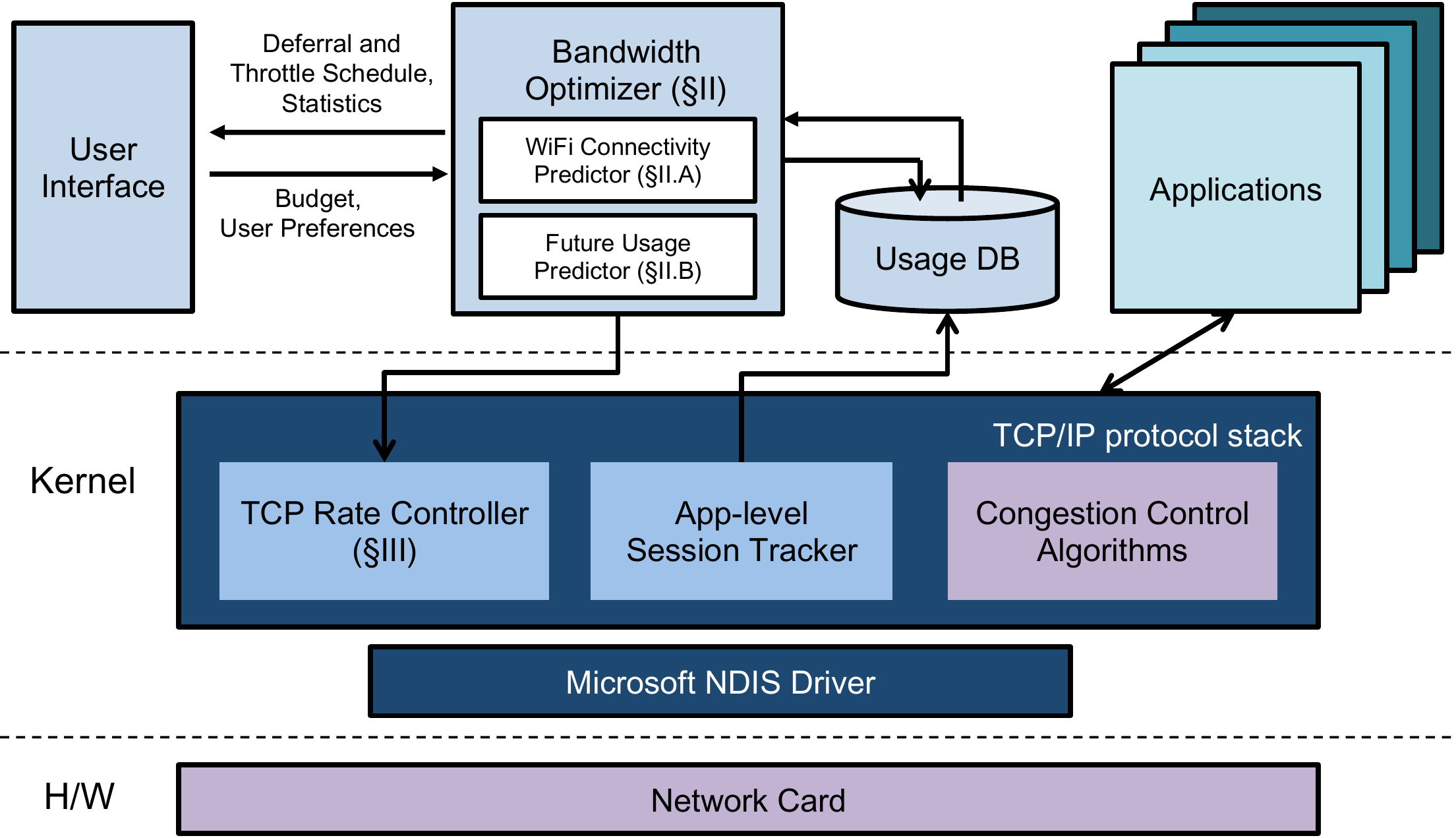}\label{fig:architecture}} \\
\vspace{-0.1in}
\subfloat[Control schematic and main modules.]{\includegraphics[width = 0.4\textwidth]{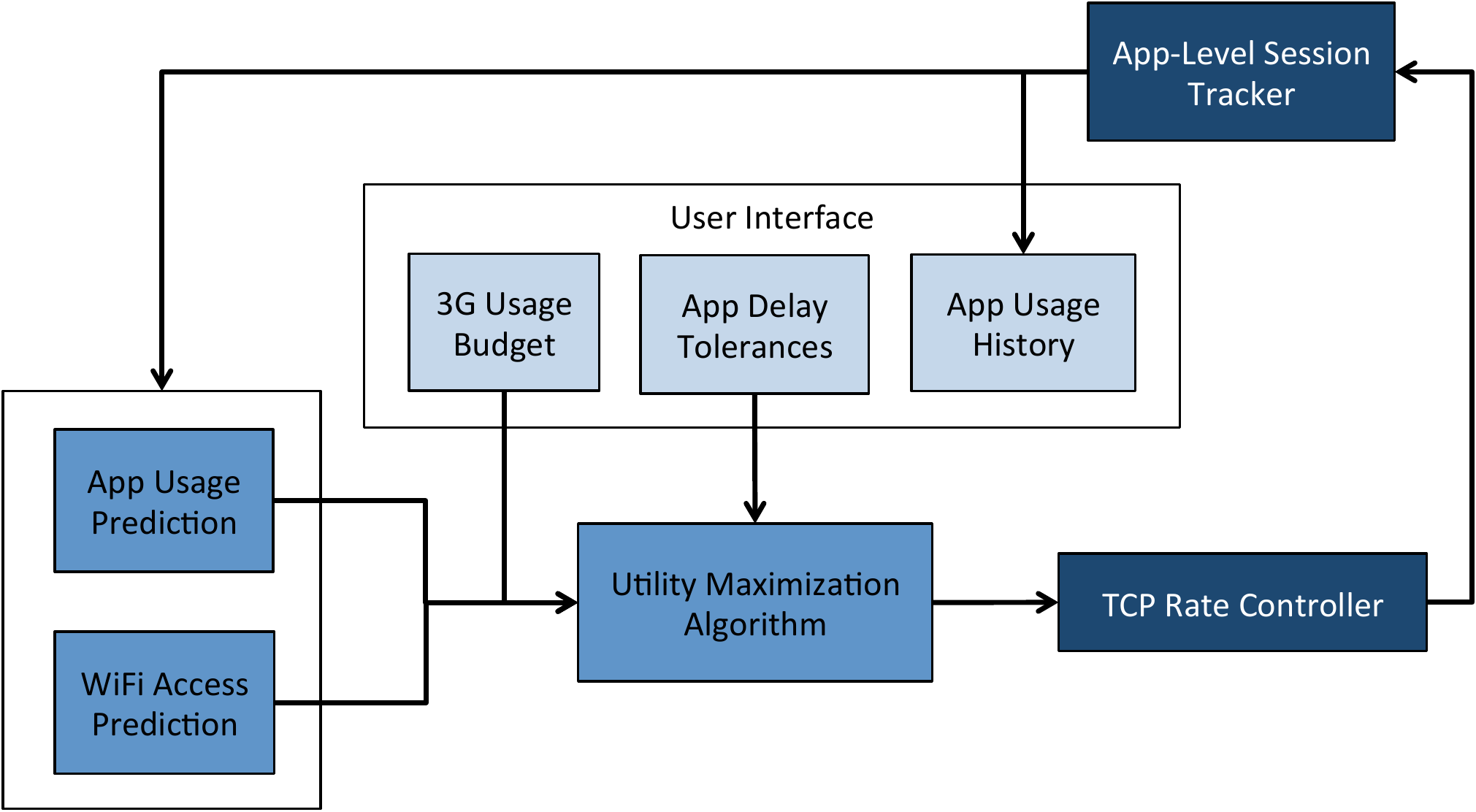}\label{fig:control}} \\\caption{Overview of AMUSE's components.}
\label{fig:overview}
\vspace{-0.12in}
\end{figure}

Since 99.7\% of mobile traffic flows are TCP, we enforce the Bandwidth Optimizer's 3G bandwidth allocations and offloading decisions with a TCP rate controller on end-user devices \cite{rahmati2011mobile}.  To do so, the controller modifies the TCP advertisement window in outgoing acknowledgement (ACK) packets.  Unlike typical bandwidth throttling mechanisms, this rate control is completely specified by the \emph{end user} on a \emph{per-application} basis; thus, for example, file downloads may be delayed to wait for WiFi, while streaming videos may receive a higher 3G bandwidth and not be delayed.  The app-level session tracker measures the actual usage for each application as the rate controller enforces the Bandwidth Optimizer's decisions.  These usage data are then used to update AMUSE's prediction modules, as shown in the control flow diagram (Fig. \ref{fig:control}), and are displayed to the user on the User Interface.

In Section \ref{sec:RelatedWork}, we discuss prior works that propose functions related to components of the AMUSE system.  Section \ref{sec:allocation_algorithm} discusses the Bandwidth Optimizer in more detail, while Section \ref{sec:Implementation} gives an overview of the TCP Rate Controller's algorithm and implementation.
In Section \ref{sec:measurement}, we observe mobile users' wireless network usage pattern from the viewpoint of WiFi offloading, and find how much and which kind of applications are currently offloaded and can be offloaded more.
In Section \ref{sec:evaluation}, we evaluate AMUSE's effectiveness in improving users' experience, utilizing 3G and WiFi data gathered from 37 mobile users.  When compared with two representative offloading algorithms (on-the-spot and delayed~\cite{OffloadingCoNext2010}), we show that AMUSE increases user utility by intelligently managing the cost-throughput-delay tradeoff for heavy and light users.  Finally, we conclude the paper in Section \ref{sec:Conclusion}.

\section{Related Work}

\label{sec:RelatedWork}

Recent studies of 3G and WiFi usage traces, e.g. \cite{OffloadingCoNext2010} have showed that offloading 3G traffic to WiFi can significantly benefit mobile ISPs. Other systems have demonstrated offloading's benefits for user experience \cite{breadcrumbs,siris2013performance,Augmenting3G}; other works demonstrate that WiFi offloading can benefit both ISPs and users \cite{6567156} and even generate more revenue for ISPs \cite{joe2013offering}.

Some works have focused on incentivizing users to offload traffic to WiFi. In \cite{6089074}, the authors develop a utility and cost-based formulation to decide the 3G network load that maximizes the user's benefit and apply the decided loads using a modified SCTP implementation in Linux that stripes traffic across multiple interfaces.
Win-Coupon~\cite{WinCoupon} takes a slightly different perspective and proposes a reverse-auction scheme to incentivize users to offload their traffic so as to decrease the overall network. 
 Other works, including~\cite{Ristanovic}, consider the energy consumption when making an offloading decision. We do not consider energy in this work, but can easily incorporate the battery consumption into our proposed optimization algorithm.

Several research works have analyzed the traffic of smart devices in order to understand their user behavior \cite{falaki2010diversity}, \cite{xu2011identifying}, \cite{falaki2010first},  \cite{gember2011comparative}, \cite{maier2010first}. 
In particular, \cite{falaki2010diversity} examines users' traffic diversity, relationship to application types, interactivity, and diurnal patterns, while \cite{xu2011identifying} investigates the usage patterns of smartphone apps via network-side measurements. Our work is different from these in that we specifically focus on the potential for WiFi offloading of different applications and incorporate findings on their delay tolerances.

Unlike most offloading works, AMUSE also controls the 3G bandwidth for different applications. AMUSE is unique in using of receiver-side TCP advertisement windows to control application-specific 3G bandwidth from the user side.  While several commercial applications (e.g., \cite{NetLimiter,SoftPerfectBM,PRTG}) provide user-side application rate control, most require users to manually specify the desired rates.  AMUSE provides automated bandwidth rates and, by conforming to TCP interactions, avoids the TCP timeouts common to existing user-side rate control applications.  
Although the TCP advertisement window is normally used by the TCP receiver to inform the TCP sender of its available buffer space, other trials \cite{1177250} have used the advertisement window as a means to control the rate of applications. However, this approach has been mainly applied to the enforcement of different application priorities, rather than direct control of the application rates.
Other solutions, such as~\cite{10.1109/TC.2004.1261834}, focus on the edge gateway, rather than the end user.

\section{Bandwidth Optimizer}\label{sec:allocation_algorithm}

In this section, we describe the individual components of AMUSE's bandwidth optimization algorithm.  Our design follows two principles: 1) AMUSE's offloading decisions will be implemented in real time on arriving sessions, and 2) AMUSE must use only the data and computational resources available on the end user's device.  Thus, we require simple, yet accurate, algorithms to compute concrete offloading decisions that can be communicated directly to the TCP Rate Controller (cf. Fig. \ref{fig:control}).  In the discussion below, we first introduce practical algorithms to predict WiFi access and application-specific usage (Sections \ref{sec:wifi} and \ref{sec:usage}).  We then incorporate these predictions into a mathematical allocation framework  in Section \ref{sec:formulation} and propose a heuristic algorithm for computing AMUSE's bandwidth allocations and offloading decisions in Section \ref{sec:online}.

To consider a user's different delay tolerances on different applications, we group a user's traffic into different application types, e.g., streaming, browsing, and downloads.  For practical implementability, we assume that only the most heavily used (e.g., top five) applications are considered, and denote these collectively as a set $J$.  We suppose that the day is divided into $n$ discrete periods of time, e.g., 24 hours, and for each period, we predict both WiFi access and application usage volumes.  

Given these predictions, we (i.e., AMUSE) must decide which applications to offload when, subject to a maximum 3G usage budget.  By delaying sessions to future periods, users may gain WiFi access and the ability to offload; however, if WiFi is unavailable, the user must send these sessions over 3G, which has a finite bandwidth capacity that must be shared among the different applications.  AMUSE therefore computes a 3G bandwidth allocation when deciding whether to wait for WiFi.  Following the first principle above, we formulate this decision as a multiple choice knapsack problem, and propose a heuristic solution algorithm. 

\subsection{Predicting WiFi Connectivity}\label{sec:wifi}


Since WiFi availability is heavily location-dependent, we predict the probabilities of WiFi access by combining user location prediction with the probabilities of WiFi access at different locations.  We define a ``location'' to be an area with WiFi coverage (e.g., a user's home).  To improve our prediction algorithm's accuracy, we consider the \emph{functional} availability of WiFi at different locations: while WiFi is always \emph{physically} available at a given location, the user may not access WiFi every time that she is there.  For instance, a user may sometimes connect to WiFi at her local Starbucks, but may walk by on weekdays without initiating a connection.  These access probabilities also depend on time: \revise{a user might stop at Starbucks in the morning but not in the evening}.  We use a training set of empirical WiFi access data to estimate these time-dependent WiFi access probabilities at each location, and modify them as we collect more access data.\footnote{One may refine these calculations by using only weekday or only weekend data, as user mobility will likely differ on weekdays and weekends.}  For a location $l$, we denote the probability of WiFi access during period $k$ as $v_k(l)$.  We use $L_k$ to denote the set of observed locations in period $k$.

Given the time- and location-specific probabilities $v_k(l)$, we then predict \emph{overall} WiFi access by incorporating predictions of users' future locations.  We define $w_k$ to be the overall WiFi probability in period $k$, \revise{i.e., the expected WiFi probability, considering the probabilities of all possible locations in period $k$ and the WiFi access probability in period $k$ for each location.}
We use a second-order Markov chain for the location prediction, which has been shown to be highly accurate \cite{song2006evaluating}.   Algorithm \ref{alg:wifi} summarizes the calculation of overall WiFi access probabilities.  We use the notation $p^{k + 2}_{l}(l_kl_{k + 1})$ to denote the probability that a user is at location $l\in L_{k + 2}$ during period $k + 2$, given his locations $l_k$ in period $k$ and $l_{k + 1}$ in period $k + 1$.  To calculate these $p^k$, we define $N^{k}(s)$ as the number of times that $s$ is observed, where $s$ is a sequence of locations that ends in period $k$; the observed location in each period $k$ is denoted by $\lambda_k$.  
\revise{We update the $p^k$ values using the empirical probabilities of a user being at different locations during period $k$}.

\subsection{Predicting Future Usage}\label{sec:usage}

At the beginning of each day, we use previous data to predict the size $s_j(k)$ of each application type $j\in J$'s usage in each period $k$.  To accommodate the dependence of session size on the amount of bandwidth allocated, our definitions of session ``size'' depend on the application: for fixed-volume application sessions such as downloads, in which the volume (MB) does not depend on the available bandwidth, we define the session size as its volume.  For fixed-time sessions such as streaming, in which the volume does depend on the bandwidth, we define the size as the time to complete.  We use $J_v$ to denote the set of fixed-volume and $J_t$ the set of fixed-time application types.  We stress that our prediction algorithms do not depend on the definition of session size; they rely only on users' consistency from day to day. 
\begin{algorithm}[t]
\begin{scriptsize}
\DontPrintSemicolon
\If{$i = 1$}{
	\For{$k \leftarrow 1$ \KwTo $n$}{
		$w_k \leftarrow \sum_{l\in L_k} v_k(l)\frac{N^k(l)}{N}$, $N$ is the number of days of data. \tcp{Calculate WiFi probabilities for the next $n$ periods.}
	}
}
\If{$i > 1$}{
	\For{$k \leftarrow 2$ \KwTo $n$}{
		\ForAll{$l\in L_k$, $l_{k-1}\in L_{k-1}$, $l_{k - 2}\in L_{k - 2}$}{
			\If{$N^{k - 1}(l_{k - 2}l_{k - 1}) > 0$}{
				$p_l^{k}(l_{k - 2}l_{k - 1}) \leftarrow \frac{N^{k}(l_{k - 2}l_{k - 1}l)}{N^{k - 1}(l_{k - 2}l_{k - 1})}$\;
			} \Else {
				$p_l^{k}(l_{k - 2}l_{k - 1}) \leftarrow \frac{N^{k}(l_{k - 1}l)}{N^{k - 1}(l_{k - 1})}$\;
			}
		}
		$w_k \leftarrow \sum_{l\in L_{k}} p_l^k(\lambda_{k - 2}\lambda_{k - 1})v_k(l)$\;
	}
}
\caption{Computation of WiFi access probabilities over the rest of the day in period $i$.\label{alg:wifi}}
\end{scriptsize}
\end{algorithm}
We estimate the future usage $s_j(k)$ by taking a moving average of the observed usage sizes $\sigma_j(k)$ of application $j$ in period $k$ over some fixed number of days.\footnote{Other prediction methods (e.g., ARIMA) can be substituted for a moving average without affecting the overall structure of our system.}

In updating our usage estimates, we modify the moving-average calculation to take into account our deferral recommendations\revise{; we wish to predict the amount of future usage without the deferrals recommended by our algorithm}.  Since a user may delay application usage to another time in order to offload it to WiFi, we ``shift the usage back'' in order to evaluate and detect changes in the underlying usage pattern over the day.  We perform these adjustments if the observed usage size for application $j$ in period $i$ is much less than the predicted $s_j(i)$, i.e., the user has shifted her usage of application $j$ from period $i$.  To account for the uncertainty in our predictions, we suppose that the actual usage deferred to period $k$ from each period $i$ is proportional to the predicted usage deferred;\footnote{\revise{We assume that at the time of deferring an application, we cannot know how much traffic the user will defer. For many applications such as web browsing and social networking, it is hard to know the exact traffic amount they will use in advance, since their contents can be dynamically selected or created by the user.}} this assumption ensures that we do not calculate that more usage was deferred to period $k$ than the actual usage observed in that period. Thus, for each application $j\in J$ and period $i< k$, we adjust the observed usage $\sigma_j(i)$ and $\sigma_j(k)$ by
\begin{align*}
\sigma_j(i) &\leftarrow \sigma_j(i) + \sum_{k = i + 1}^n\frac{c_i^j(k)s_j(i)\sigma_j(k)}{s_j(k) + \sum_{l = 1}^{k-1}c_l^j(k)s_j(l)},\\
\sigma_j(k) &\leftarrow \sigma_j(k) - \sum_{i = 1}^{k - 1}\frac{c_i^j(k)s_j(i)\sigma_j(k)}{s_j(k) + \sum_{l = 1}^{k-1}c_l^j(k)s_j(l)}.
\end{align*}
Here $c_i^j(k)$ is an indicator variable taking the value 1 if application $j$ is deferred from period $i$ to period $k$ and 0 otherwise. \revise{The first expression adjusts the observed usage for application $j$ in period $i$ by adding the traffic amount deferred to later time periods, while the second expression adjusts by subtracting the traffic amounts deferred from previous times. This method is approximate, but we expect that it will be helpful for predicting the future traffic amounts from observed traffic.}   


\subsection{User Utility Maximization}\label{sec:formulation}

In this section, we formulate the user's offloading decision problem, assuming the future WiFi probabilities $w_k$ and usage $s_j(k)$ are known.  In the discussion below, the phrase ``originally in period $i$'' indicates that the application session(s) under consideration are completed in period $i$ if they are not deferred to a future period.

\subsubsection{Utility Functions}

To mathematically formulate the user's offloading decision problem, we need a concrete measure of the user's tradeoffs between cost, throughput, and delay.  Thus, for a given application type $j$ in period $i$, we derive expressions for users' \emph{utility} of completing those application sessions over 3G and over WiFi.  This utility is determined by the per-volume price $p$ of 3G, the amount of time $t$ the session is deferred, the bandwidth speed $r$ at which the session is completed, and the size $s$ of the session.  We use $U_j(p, t, r, s)$ to denote the utility of application $j\in J$.

\reviseAgain{Though many functions could be used as the $U_j$, we note that these should be decreasing in $p$ and $t$ (price and time deferred) and increasing in $r$ (bandwdith). \revise{For fixed-volume applications, we suppose that the utility is decreasing in $s$, since a larger size indicates more time to complete the session.}
We use the economic principle of diminishing marginal utility to argue that as $r$ becomes larger or $t$ becomes smaller, users' marginal utility from $r$ should decrease, and the marginal utility from $t$ should increase.}  For simplicity, we take the units of $t$ to be the number of periods deferred, and do not consider sessions' timing \emph{within} the period to which they are deferred.  Since different users will have different tradeoffs between cost, quality, and delay, we suppose that the $U_j$ functions take the same form, but have different parameters that depend on the particular application and user.

The above guidelines still leave many possible utility functions.  To narrow these down, we conducted an online survey of over 100 users, primarily students, faculty and staff from U.S. universities.  For each application in Table \ref{tab:param}, we gave participants the cost to complete one application session over 3G, as well as the speed of WiFi relative to 3G.  We then asked the participants how long they would wait for WiFi access instead of immediately completing the session over 3G; for each question, we offered five options for the maximum amount of time participants were willing to wait, ranging from ``I won't wait'' to ``as long as necessary.''\footnote{The survey questions are available in \cite{tr}.}

We find that our survey data provides a good fit with the functional form
\begin{equation}
\begin{cases}
U_j(p, t, r, s) = C_j\exp\left(-\nu + r\nu - \mu t\right) - \eta prs &j\in J_t \\
U_j(p, t, r, s) = C_j\exp\left(-(s/r)\nu - \mu t\right) - \eta ps &j\in J_v,
\end{cases}
\label{eq:util}
\end{equation}
where $U_j$ denotes the parameterized utility function for application types $j$; $prs$ for $j\in J_t$ and $ps$ for $j\in J_v$ denote the cost of each session; and $C_j$, $\mu$, $\nu$, and $\eta$ are nonnegative parameters that depend on $j$.  These functions satisfy several desirable properties: for example, the constants $C_j$ allow for different user priorities for different types of sessions (e.g. a user intrinsically derives more utility from certain applications, even with zero delay or time to completion).\footnote{The $-\nu$ in the exponential for $j\in J_t$ is included for normalization: with maximum bandwidth 1 and no delay, we then have $U_j = C_j$.}  For $j\in J_v$, the $s/r$ term in the exponential represents the time to completion, while for $j\in J_t$, the bandwidth $r$ represents the quality of the streaming video. 


Table \ref{tab:param} shows the parameter values calculated for each session type. 
To estimate these parameter values, we used the probability that a user would not wait for WiFi as the utility function value, with $b$, $t$, $r$ and $s$ measured relative to their values for WiFi.  We assume a negligible cost term for low-volume (e.g., email) sessions.  We then used nonlinear curve-fitting methods to calculate the utility function parameters, and found a small average squared-error of 0.05 for each survey question, upon comparing the actual answers with our estimates.  

We see that the $C_j$ coefficients roughly match our expectations, with email the most important and social networking (photo uploads) the least important applications. The parameter $\mu$ represents the amount of time that a user will wait to start an application, e.g., in anticipation of WiFi access or higher 3G speeds: it is largest (i.e., users are least willing to wait) for browsing and email. The importance of available
throughput is parameterized by $\nu$, and is highest for video and lowest for social networking. 
\begin{table}
\renewcommand{\arraystretch}{1.1}
\caption{Estimated parameters for the utility function (\ref{eq:util}).}
\label{tab:param}
\vspace{-0.1in}
\centering
\begin{tabular}{|r||cccc|}
\hline
& $C$ & $\mu$ & $\nu$ & $\eta$ \\ \hline
Email & $0.9848$ & $0.1527$ & $0.1527$ & assumed 0 \\
Browsing & $0.6865$ & $0.3269$ & $0.0263$ & assumed 0 \\
Video & $0.9399$ & $0.0144$ & $4.3785$ & $0.0986$ \\
Social netw. & $0.4738$ & $0.006$ & $0.006$ & $0.0986$ \\
Downloads & $0.6737$ & $0.0097$ & $0.0097$ & $0.0986$ \\ \hline
\end{tabular}
\vspace{-0.1in}
\end{table}

\subsubsection{Users' Optimization Problem}\label{sec:optProblem}

We now use the utility functions (\ref{eq:util}) to formulate the user's optimization problem.  To represent possible 3G and WiFi bandwidth speeds, we normalize the volume units so that the fixed per-second WiFi speed equals 1.  The 3G speed $\gamma$ is chosen from a finite subset of possibilities $\Gamma$\revise{; generally, all $\gamma < 1$ since 3G speeds are usually slower than WiFi, though for LTE networks we may have $\gamma > 1$}.  For each $\gamma\in \Gamma$ and period $k\geq i$, we define the indicator variables $c_i^j(k,\gamma)$ to be 1 if a session of type $j$ is deferred from period $i$ to period $k$ and assigned 3G speed $\gamma$, and 0 otherwise. \revise{Note that $\gamma$ is always chosen for each delayed period $k$; this speed $\gamma$ is then used if WiFi is not available in period $k$. If the probability of WiFi availablity is 100\%, any value of $\gamma$ can be chosen without affecting the user's expected utility from WiFi in period $k$. This utility,} for a session of type $j$ originally in period $i$, is then
\begin{equation*}
\left(\sum_{\gamma\in\Gamma}c_i^j(k,\gamma)\right)w_kU_j\left(0, k - i, 1, s_j(i)\right),
\end{equation*}
while the expected utility from 3G in period $k$ is
\begin{equation*}
\displaystyle\sum_{\gamma\in\Gamma}c_i^j(k,\gamma)(1 - w_k)U_j\left(p, k - i, \gamma, s_j(i)\right).
\end{equation*}
The user wishes to maximize the sum of these utilities over all (original) periods $i$ and application types $j$:
\begin{align}
\max_{c_i^j(k,\gamma)}\;\sum_{i = 1}^n\Bigg[&\sum_{j\in J}\Bigg(\sum_{k\geq i}\Bigg(\sum_{\gamma\in\Gamma}\Big(w_kU_j\big(0, k - i, 1, s_j(i)\big) +\nonumber \\
&(1 - w_k)U_j\big(p, k - i, \gamma, s_j(i)\big)\Big)c_i^j(k,\gamma)\Bigg)\Bigg)\Bigg] \label{eq:obj} \\
{\rm s.t.}\;\sum_{k\geq i} \sum_{\gamma\in\Gamma}& c_i^j(k,\gamma) = 1;\;c_i^j(k,\gamma) \in\left\{0,1\right\}, \label{eq:var}
\end{align}
where (\ref{eq:var}) ensures that each application $j$ in period $i$ is deferred to only one period $k$ (we may have $k = i$), with 3G speed $\gamma$.  This optimization is performed subject to two constraints: a budget constraint on expected 3G usage, and capacity constraints on the 3G bandwidth in each period.

We assume that the user specifies a maximum monthly budget $\overline{B}$ for 3G usage.  We then calculate a \emph{daily budget} $B$, taking into account both the number of days remaining in the month (denoted by $m$) and the amount of budget $B_r$ that has not yet been spent.  To allow the user some flexibility, we multiply the average usage $B_r/m$ by the factor $\exp\left(1 - m^{-1}\right)$, which equals 1 only if $m = 1$: at the end of the month, the user cannot exceed the remaining budget. The daily budget $B$ is then defined as $B_r \exp\left(1 - m^{-1}\right)/m$, and the budget constraint is
\begin{align}
p\sum_{i = 1}^n\Bigg[&\sum_{j\in J_v}\left(\sum_{k\geq i}\sum_{\gamma\in\Gamma} c_i^j(k,\gamma)(1 - w_k)s_j(i)\right) + \nonumber \\
&\sum_{j\in J_t}\left(\sum_{k\geq i}\sum_{\gamma\in\Gamma} c_i^j(k,\gamma)(1 - w_k)\gamma s_j(i)\right)\Bigg]\leq B,
\label{eq:budget}
\end{align}
The 3G bandwidth capacity constraints ensure that the sum of the bandwidth allocated to each application in a given period does not exceed the fixed maximum bandwidth, which we denote as $\beta$.  Mathematically, this constraint is
\begin{equation}
(1 - w_l)\sum_{i\leq l}\sum_{j\in J}\sum_{\gamma\in\Gamma} c_i^j(l,\gamma)\gamma\leq (1 - w_l)\beta.
\label{eq:capacity}
\end{equation}
We include a $1 - w_l$ term on each side so that if $w_l = 1$ and all sessions complete over WiFi, any 3G speed $\gamma$ may be chosen.  Putting (\ref{eq:obj}-\ref{eq:capacity}) together, we obtain the optimization problem
\begin{align}
\max_{c_i^j(k,\gamma)}\;\sum_{i = 1}^n&\Bigg[\sum_{j\in J}\Bigg(\sum_{k\geq i}\Bigg(\sum_{\gamma\in\Gamma}\Big(w_kU_j\big(0, k - i, 1, s_j(i)\big) +\nonumber \\
&(1 - w_k)U_j\big(p, k - i, \gamma, s_j(i)\big)\Big)c_i^j(k,\gamma)\Bigg)\Bigg)\Bigg] \label{eq:newobj} \\
{\rm s.t.}\;p\sum_{i = 1}^n\Bigg[&\sum_{j\in J_v}\left(\sum_{k\geq i}\sum_{\gamma\in\Gamma} c_j^i(k,\gamma)(1 - w_k)s_j(i)\right) + \nonumber \\
&\sum_{j\in J_t}\left(\sum_{k\geq i}\sum_{\gamma\in\Gamma} c_j^i(k,\gamma)(1 - w_k)\gamma s_j(i)\right)\Bigg]\leq B \label{eq:newbudget} \\
(1 - w_l&)\sum_{i\leq l}\sum_{j\in J}\sum_{\gamma\in\Gamma} c_j^i(l,\gamma)\gamma\leq (1 - w_l)\beta\,\forall\,l \label{eq:newcapacity} \\
\sum_{k\geq i} \sum_{\gamma\in\Gamma} &c_j^i(k,\gamma) = 1\,\forall\,j\in J;\;i = 1,2,\ldots,n \label{eq:newvar} \\
c_j^i(k,\gamma)&\in\left\{0,1\right\}. \nonumber
\end{align}
\revise{Note that the problem's decision variables $c_i^j(k,\gamma)$ for all time periods $i$ and applications $j$ represent the schedule for application deferrals and 3G rates to be used if the WiFi is not available at the scheduled times. Other variables such as $w_k$ and $s_j(i)$ are calculated by empirical data as explained in Sections \ref{sec:wifi} and \ref{sec:usage}.} We can view the constraints (\ref{eq:newvar}) as choosing exactly one item from a knapsack, where each $(i,j)$ pair for $i = 1,2,\ldots,n$ and $j\in J$ is associated with a knapsack consisting of items indexed by the variables $k\geq i$ and $\gamma\in\Gamma$. With this interpretation, (\ref{eq:newobj}-\ref{eq:newvar}) can be seen as a multidimensional, multiple choice knapsack problem.\reviseAgain{
In the Appendix~\ref{sec:extension}, we show that we can easily extend this formulation so that the varying WiFi capacity is considered and the user chooses which network to use depending on the utility values of WiFi and 3G.}

\subsection{Online Algorithm}\label{sec:online}

In this section, we present an online algorithm to solve the optimization problem (\ref{eq:newobj}-\ref{eq:newvar}).  At the beginning of each day, the user computes an initial solution, given estimates of the $w_k$ and $s_j(k)$.  As the $w_k$ estimates and known usage amounts are updated over the day, this solution is refined.

While various algorithms exist to compute solutions of the knapsack problem (\ref{eq:newobj}-\ref{eq:newvar}) to different degrees of accuracy, we use a Lagrange-multiplier based solution \cite{moser1997algorithm} that has relatively small computational overhead and generally returns good approximations to the optimum.\footnote{Since our estimates of WiFi access and future usage are already approximations, even an exact solution to the optimization (\ref{eq:newobj}-\ref{eq:newvar}) will be an approximation of the ``true'' optimum.}  Given a feasible solution, the algorithm improves the solution while maintaining its feasibility, allowing us to update previously computed solutions over the day.

This Lagrange multiplier algorithm starts from a solution that maximizes (\ref{eq:newobj}) without considering the constraints  (\ref{eq:newbudget}-\ref{eq:newvar}).
The solution is then adjusted so that all constraints are satisfied, beginning with the ``most violated'' (i.e., the constraint with largest Lagrange multiplier).  This process repeats until no constraints are violated, and the solution can then be improved by adjusting the solution one variable at a time, so as to most increase the objective value while still not violating the constraints.\footnote{\label{footnote_constraint}If the constraints are especially tight, the Lagrange multiplier algorithm may not yield a solution.  While in practice such a situation is unlikely to occur, we can easily recover from this failure by taking as the initial allocation the worst-case scenario, in which all sessions are given the lowest possible bandwidth; we assume that this is a feasible solution. \revise{If only the bandwidth constraint is violated, we can simply scale down the 3G bandwidths assigned to different applications.}}


As the user consumes data over the day, we update both the remaining daily budget $B$ and our predictions of future WiFi connectivity $\left\{w_k\right\}$.  The new optimization problem over the remainder of the day can then be solved by taking the existing solution as the initial point of our Lagrange multiplier algorithm.  This solution may well be feasible: the 3G capacity constraints do not change unless WiFi becomes definitely available in some period ($w_k\rightarrow 1$), in which case that period $k$'s capacity constraint is  removed.  Thus, if the existing solution satisfies the new budget constraint, we can skip directly to the ``solution improvement'' step, significantly reducing the computational overhead.  Algorithm \ref{alg:everything} presents this full online algorithm, along with the WiFi and app usage predictions (Sections \ref{sec:wifi} and \ref{sec:usage}).

\begin{algorithm}[t]
\begin{scriptsize}
\DontPrintSemicolon
$i \leftarrow 1$ \tcp{The current period is denoted by $i$.}
$B\leftarrow \left(B_r/m\right)\exp\left(1 - m^{-1}\right)$ \tcp{Calculate the budget for the day.}
Calculate WiFi probabilities using Algorithm \ref{alg:wifi} with $i = 1$.\;
Calculate predicted usage over all $n$ periods using a moving average.\;
Allocate bandwidth by approximately solving (\ref{eq:newobj}-\ref{eq:newvar}).\;
\For{$i\leftarrow 2$ \KwTo $n$}{
	$B\leftarrow B - S_{i - 1}$ \tcp{Remaining daily budget, given the spending $S_{i - 1}$ in period $i - 1$.}
	Update WiFi probabilities using Algorithm \ref{alg:wifi}.\;
	Update bandwidth allocations by re-solving (\ref{eq:newobj}-\ref{eq:newvar}) for the remaining $n - i + 1$ periods.\;
}
\caption{Bandwidth allocation over a day.\label{alg:everything}}
\end{scriptsize}
\end{algorithm}

\section{Implementation}

\label{sec:Implementation}

We implemented an AMUSE prototype on Windows 7 tablets with the system architecture shown in Fig. \ref{fig:architecture}. 
We used the Windows Filtering Platform (WFP)~\cite{WFPCalloutDriver} to track application usage and implement a user-side TCP rate control algorithm to control each application's download rate. 

\begin{figure}[t]
\centering 
\centering \includegraphics[scale=0.215]{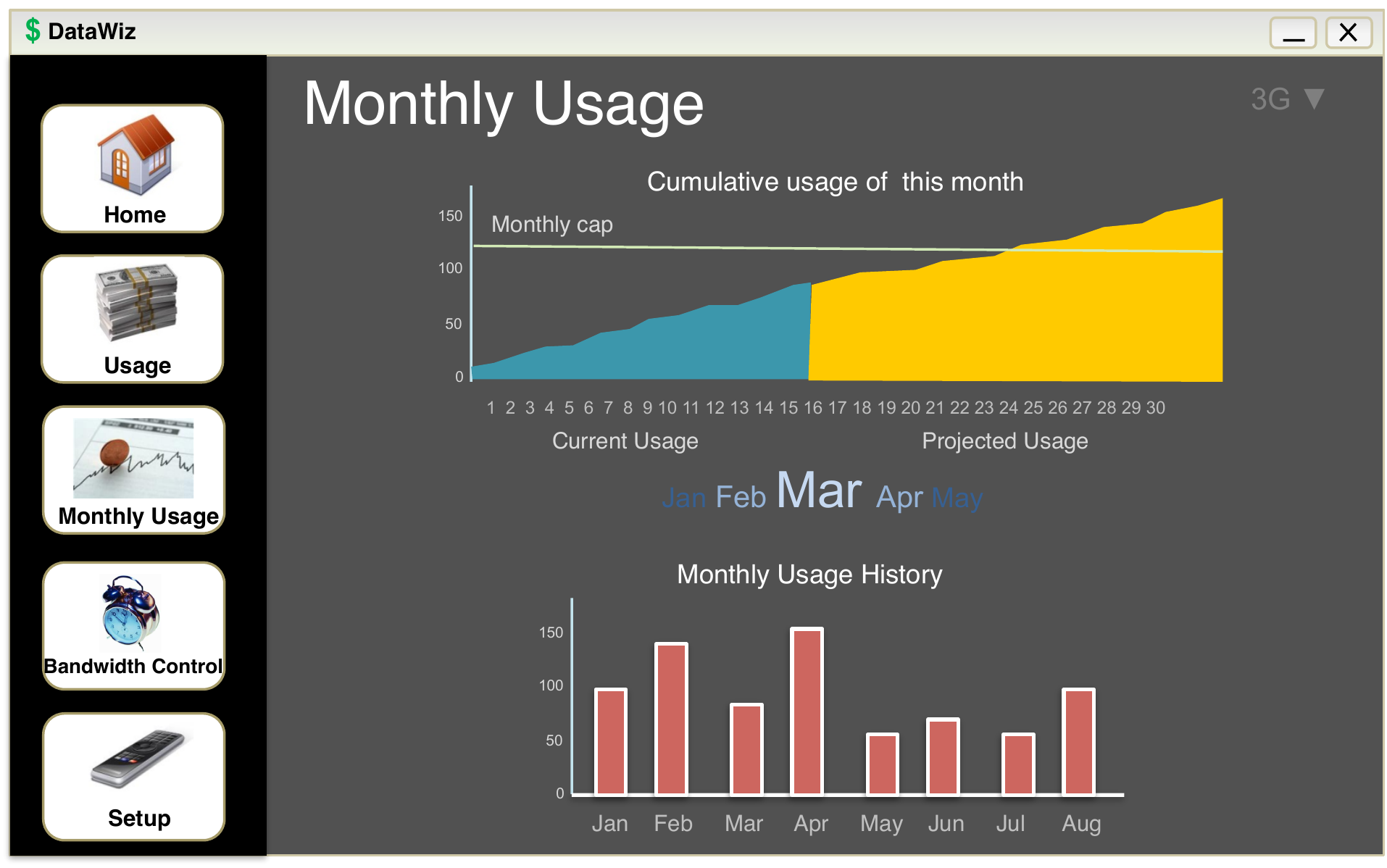}\hspace{0.1in}
\includegraphics[scale=0.215]{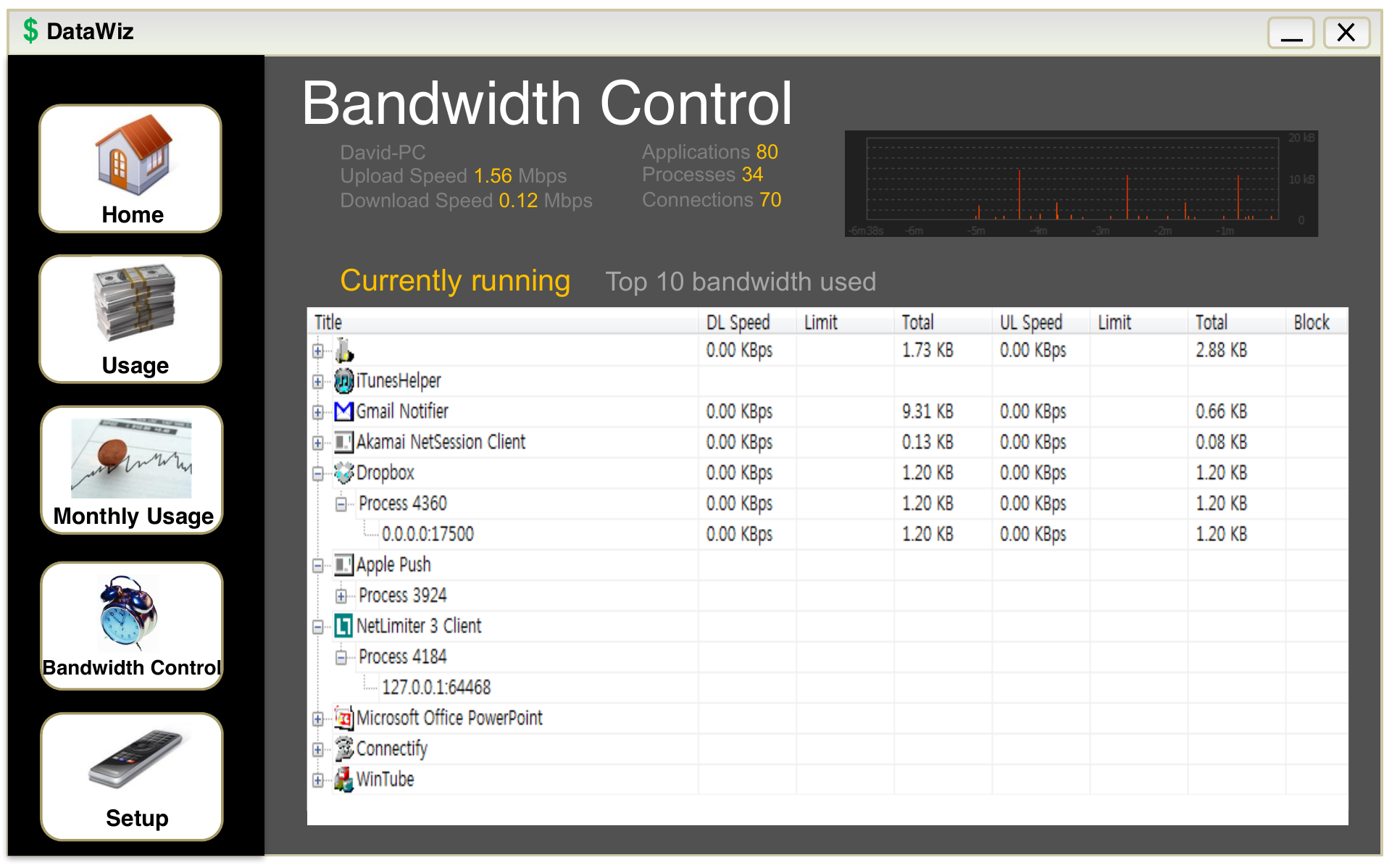} 
\vspace{-0.1in}
\caption{Screenshots of the AMUSE prototype. Users can view their monthly usage and set bandwidth rates for individual applications.}
\label{fig:amuse_screenshot}
\vspace{-0.1in}
\end{figure}

The AMUSE prototype displays both total usage and the usage of individual applications on a daily, weekly, and monthly basis, as well as the current upload and download rates. 
We also provide user interfaces from which the user can, if he so chooses, set the download rate of each application and configure his billing starting date and data plan (e.g., 2GB per month). Figure \ref{fig:amuse_screenshot} shows screenshots of these features.

We next describe the details of our receiver-side TCP rate control algorithm, followed by experimental data verifying its efficacy.

\subsection{Receiver-Side TCP Rate Control }
\label{sec:BWControl}

\revise{Algorithm~\ref{alg:everything} decides the 3G rate that will be used if WiFi is not available during the scheduled time period. In this section, we introduce a TCP rate control algorithm that applies the decided 3G rate to the receiver-side TCP connection, enforcing the result of Algorithm~\ref{alg:everything}. We devised a receiver-side TCP rate control algorithm since TCP accounts for most Internet traffic~\cite{rahmati2011mobile}, and sender-side TCP rate control can be easily implemented (e.g. using $tc$ in the Linux system).}

A TCP sender adjusts a session's rate based on its congestion window size ($cwnd$).  The ACK packets from the receiver act as a feedback to the sender on how much has been sent and how much more can be sent to the receiver. 
We use this ACK clocking to shape the incoming/downloading rates of TCP traffic, by modifying the TCP advertisement window size ($rcv\_wnd$) field in each ACK packet using the WFP driver. 
The idea behind this approach is that a TCP sender cannot send more than ${\rm min} \left(cwnd, rcv\_wnd\right)$.  While one could instead shape the TCP traffic rates by adjusting the round-trip time (RTT) of each flow (i.e., stretching each ACK packet), this latter approach increases the overall response time and renders some interactive or video TCP applications useless. Modifying the advertisement window size does not increase the RTT of each flow, making it suitable for all TCP applications.


Unlike current traffic control tools, our proposed control mechanism does not forcibly drop incoming packets, a measure that can induce such undesirable side effects as frequent TCP timeouts. The principle behind our mechanism is as follows:
\textit{we increase the size of the advertisement window if the traffic rate recently received is smaller than the target bandwidth,
and decrease it if the traffic rate recently received is larger than the target bandwidth}.
With this approach, we can implement the bandwidth control
entirely at the TCP receiver. The sender is not modified, but it will react to the advertisement window from the receiver according to TCP flow control.\footnote{In order to not hurt a user's response time with short-lived TCP flows, the algorithm only runs after 5 secs, during which these short-lived flows can complete their transfers.}
\begin{algorithm}[t]
\begin{scriptsize}
\DontPrintSemicolon
Initialization:\;
$~~~target\_BW \leftarrow$ \tcp{ Desired bandwidth (bytes/sec)}
$~~~min\_adv\_win \leftarrow 512 ~(bytes)$\;
$~~~adv\_win \leftarrow min\_adv\_win$\;
$~~~last\_check\_time \leftarrow current\_time ~(sec)$\;
$~~~check\_period \leftarrow 0.2 ~(sec)$\;
$~~~bytes \leftarrow 0 ~(bytes)$\; \tcp{accumulated received bytes for current period}
$~~~\alpha \leftarrow 0.5$ \tcp{smoothing factor}
For each TCP data and ACK packet:\;
\Begin{
	\If {a data packet is received}{
		$bytes \leftarrow bytes + packet\_len$\;
		\If {$current\_time-last\_check\_time>check\_period$}{
			$interval \leftarrow current\_time - last\_check\_time$\;
			$throughput \leftarrow bytes / interval$\;
			$inc \leftarrow adv\_win * \frac{target\_BW - throughput}{target\_BW} * \alpha$\;
			$adv\_win \leftarrow adv\_win + inc$\;
			\If {$adv\_win>rcv\_buf\_size$}{
				$adv\_win \leftarrow rcv\_buf\_size$\;
			}
			\ElseIf {$adv\_win<min\_adv\_win$}{
				$adv\_win \leftarrow min\_adv\_win$\;
			}
			$last\_check\_time \leftarrow current\_time$\;
			$bytes \leftarrow 0$\;
		}
	}
	\If {an ACK packet is ready to be sent}{
		set the advertisement window of the ACK to $adv\_win$\;
	}
}
\caption{Receiver-side TCP rate control.~\label{alg:advWinControl}}
\vspace{-0.05in}
\end{scriptsize}
\end{algorithm}

Algorithm \ref{alg:advWinControl} presents the pseudo code of our implementation.
The algorithm first initializes the $adv\_wnd$ to the default value ($min\_adv\_win$) when the connection is set up,
and periodically calculates the traffic throughput for each application in each period.\footnote{We set this value
to 200 msec. We found this value works well in various settings after comprehensive experiments.}
The throughput is obtained by dividing the received bytes ($bytes$) by the interval length ($interval$). If the throughput for a given period is smaller than the target bandwidth ($target\_BW$), we increase the advertisement window size by an amount ($inc$) proportional to the deficit throughput.
Similarly, if the throughput is larger than the target bandwidth, we decrease the size of the advertisement window
by an amount ($dec$) proportional to the surplus throughput.
Depending on the increase/decrease of the advertisement window, the TCP sender
will increase or decrease the rate of the traffic accordingly, assuming its congestion window is mostly larger than its advertisement window.
Here, we multiply the deficit/surplus bandwidth by a ratio $\alpha$, in order to reduce the
oscillation of throughput due to the drastic window size changes.
We use $\alpha = 0.5$ after experimentally determining this value's efficacy in achieving the target bandwidth in
several different environments.
We prevent the advertisement window size from moving above the maximum buffer size ($rcv\_buf\_size$) and below minimum window size ($min\_adv\_win$).

\subsection{Experimental Efficacy}
\label{sec:BWControlEval}

To verify Algorithm \ref{alg:advWinControl} in practice, we first test our receiver-side bandwidth control algorithm by running \texttt{Iperf} over Ethernet, WiFi, and 3G networks.
We used target bandwidths of 100 Kbps, 500 Kbps, and 1 Mbps;  experimental results for the three cases are shown in Table~\ref{tab:basic_bc_op}.
While the bandwidth control algorithm achieves the target rate in each of the three different networks, we observe that the rate over the Ethernet link is much closer to the target rate than the rates over WiFi and 3G: packet loss rate and link jitter are the smallest in Ethernet. 

\begin{table}[t]
\renewcommand{\arraystretch}{1.1}
\caption{Basic rate control test using Iperf. Parentheses denote the standard deviations.} \label{tab:basic_bc_op}
\vspace{-0.1in}
\centering
\begin{tabular}{|c||c|c|c|}
\hline Target rate & 100 Kbps & 500 Kbps & 1,024 Kbps \tabularnewline \hline \hline 
Ethernet & 103.8 (0.42) & 506.2 (0.42) & 1031.2 (1.81)  \tabularnewline \hline 
WiFi & 83.14 (3.63) & 459 (6.46) & 902.4 (21.67) \tabularnewline \hline 
3G & 95.28 (1.52) & 474.7 (11.86) & 896 (47.28)
\tabularnewline \hline
\end{tabular}
\end{table}

We also test the algorithm with different applications. For this experiment, we set the target bandwidth to 300 Kbps and run two applications (HTTP and FTP). 
Our results (Table~\ref{tab:application_bc_op}) show that the rates achieved are similar to the target rate. 

\begin{table}[t]
\renewcommand{\arraystretch}{1.1}
\caption{Application rate control test using HTTP and FTP. Parentheses denote the standard deviations.} \label{tab:application_bc_op}
\vspace{-0.1in}
\centering
\begin{tabular}{|c||c|c|}
\hline Application (rate) & HTTP (300 Kbps) & FTP (300 Kbps) \tabularnewline \hline \hline 
Ethernet & 297.04 (3.54) & 291.2 (4.68)\tabularnewline \hline 
WiFi & 271.12 (10.77) & 269.28 (8.27) \tabularnewline \hline 
3G & 296.16 (3.33) & 279.2 (4.43)
\tabularnewline \hline
\end{tabular}
\vspace{-0.05in}
\end{table}

\begin{figure}[t]
\vspace{-0.2in}
\centering \includegraphics[scale=0.50]{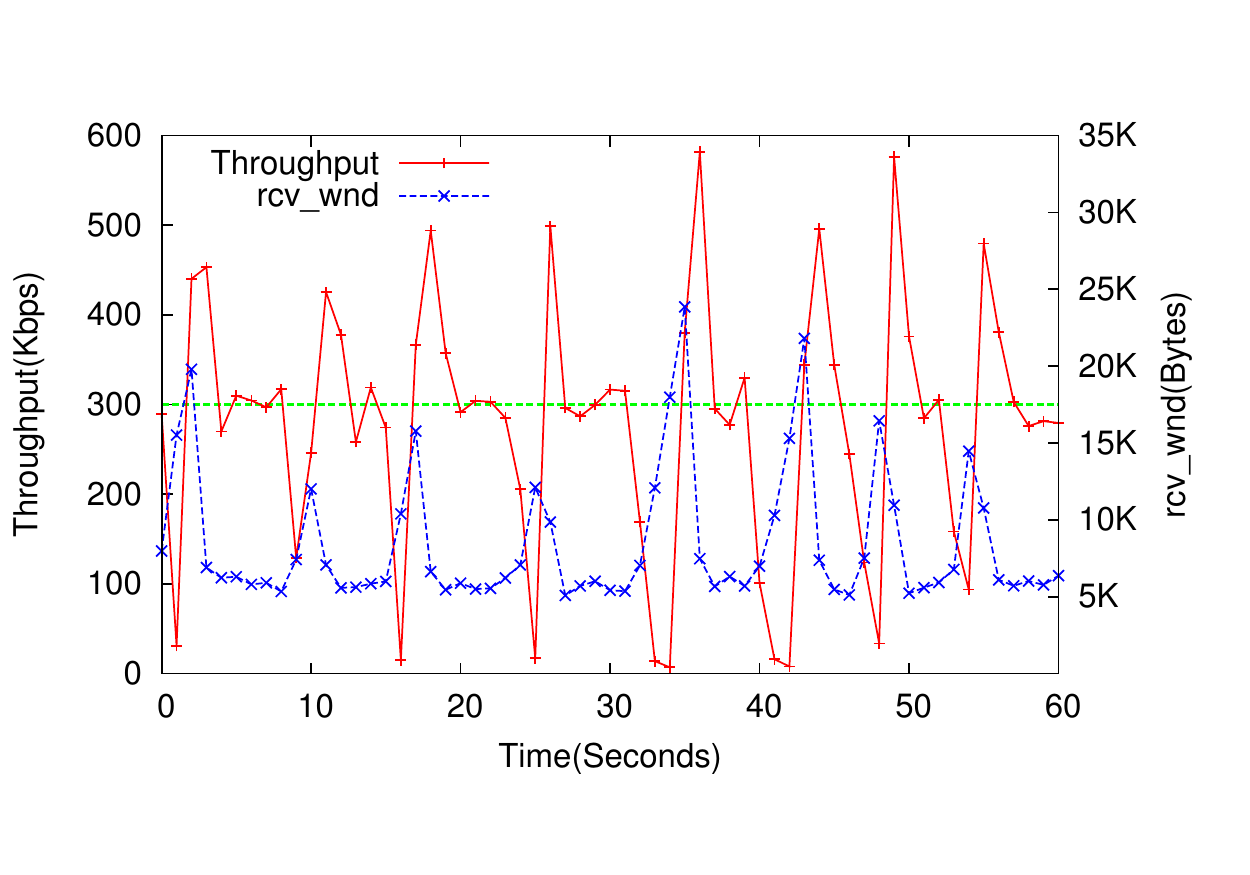} 
\vspace{-0.3in}
\caption{An example of throughput and advertisement window variations.}
\label{fig:advWin_throughput_variation}
\end{figure}

Finally, we show the time evolution of the advertisement window size $rcv\_wnd$ and the resulting downloading rate in
Figure \ref{fig:advWin_throughput_variation} for one FTP flow with a target bandwidth of 300 Kbps. The target bandwidth is represented by a straight green line.
The time evolution of $rcv\_wnd$ and the rate behave as expected: if the flow rate is smaller than the target, then the window size increases, increasing the rate after a delay. The opposite behavior is seen if the flow rate is larger than the target, but eventually both the flow rate and window size stabilize.

\section{Measurement} \label{sec:measurement}
We conduct a measurement study in order to analyze mobile users' network usage pattern from the viewpoint of WiFi offloading.
Specifically, in Section \ref{sec:application_types} we show that the application types considered in Table \ref{tab:param} comprise a large portion of users' cellular traffic. In  Section \ref{sec:offloading_practice}, we examine the degree to which these applications are already offloaded and their potential for more offloading.

\subsection{Data Collection} \label{sec:data_collection}
\begin{figure}
\centering
\includegraphics[height = 0.22\textwidth]{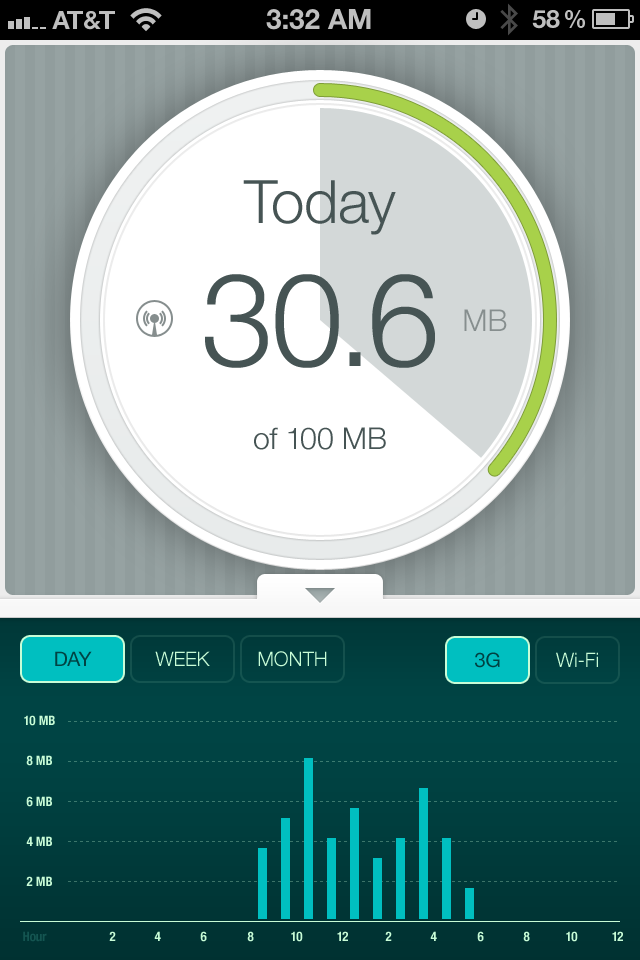}
\includegraphics[height = 0.22\textwidth]{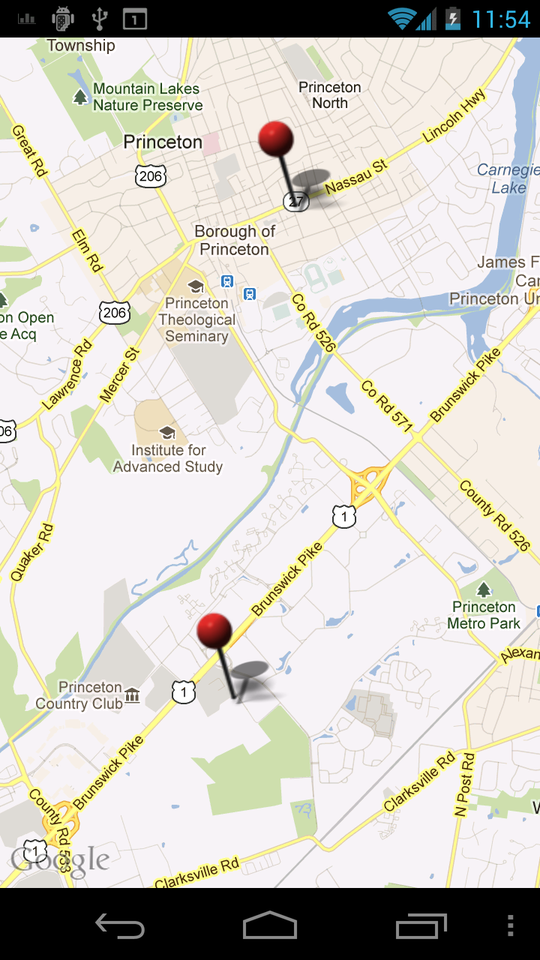}
\caption{Screenshots of the usage monitoring app.}
\label{fig:screen}
\end{figure}

To collect empirical traffic data for the measurement study in this section,
we recruited smartphone users to participate in our trial. We recorded the data by implementing a usage monitoring app and installing it on users' phones.  Figure \ref{fig:screen} shows the screenshots of the usage monitoring app, which informed users of their overall usage over different timescales and usage at different geographical locations. 

We collected application usage data from 20 Android smartphone users in Alaska for 7 days, including application package names and categories and upload and download usage amounts for each application in bytes. 
To compare AMUSE to other offloading algorithms in Section \ref{sec:evaluation}, we also collected another data set from an additional 12 Android users and 25 iPhone users in the U.S. This second dataset includes participants' 3G and WiFi usage, WiFi availability, and user locations at a ten minute granularity.\footnote{We do not collect iPhone application usage data due to iOS implementation restrictions.}

\subsection{Application types} 
\label{sec:application_types}

%

In Section \ref{sec:allocation_algorithm}, we classify user's traffic into 5 application types (i.e. Email, Browsing, Video, Social networking, Downloads). In this subsection, we verify that these application types comprise most of users' traffic by volume, indicating that AMUSE covers most cellular traffic. In Tables \ref{tab:top_wifi} and \ref{tab:top_cellular}, we list the top 15 applications for WiFi and cellular networks, respectively. 
To identify the application types, we manually searched for the package names in the Android application market and used the application descriptions there. Packages not found in the Android application market were classified using the name itself (e.g. com.android.email is classified as ``Email''). Package names that cannot be identified using these methods are designated as ``Unclassified.'' In the case of cellular network, the top five applications correspond to the application types in Table \ref{tab:param}, accounting for 54\% of the total cellular traffic. From these observations, we can conclude that our offloading mechanism can handle a large portion of mobile users' cellular traffic, demonstrating the possibility of significantly reducing the data cost.

\subsection{Offloading practice} 
\label{sec:offloading_practice}

\begin{figure}[t]
\vspace{-0.2in}
\centering
\includegraphics[scale=0.45]{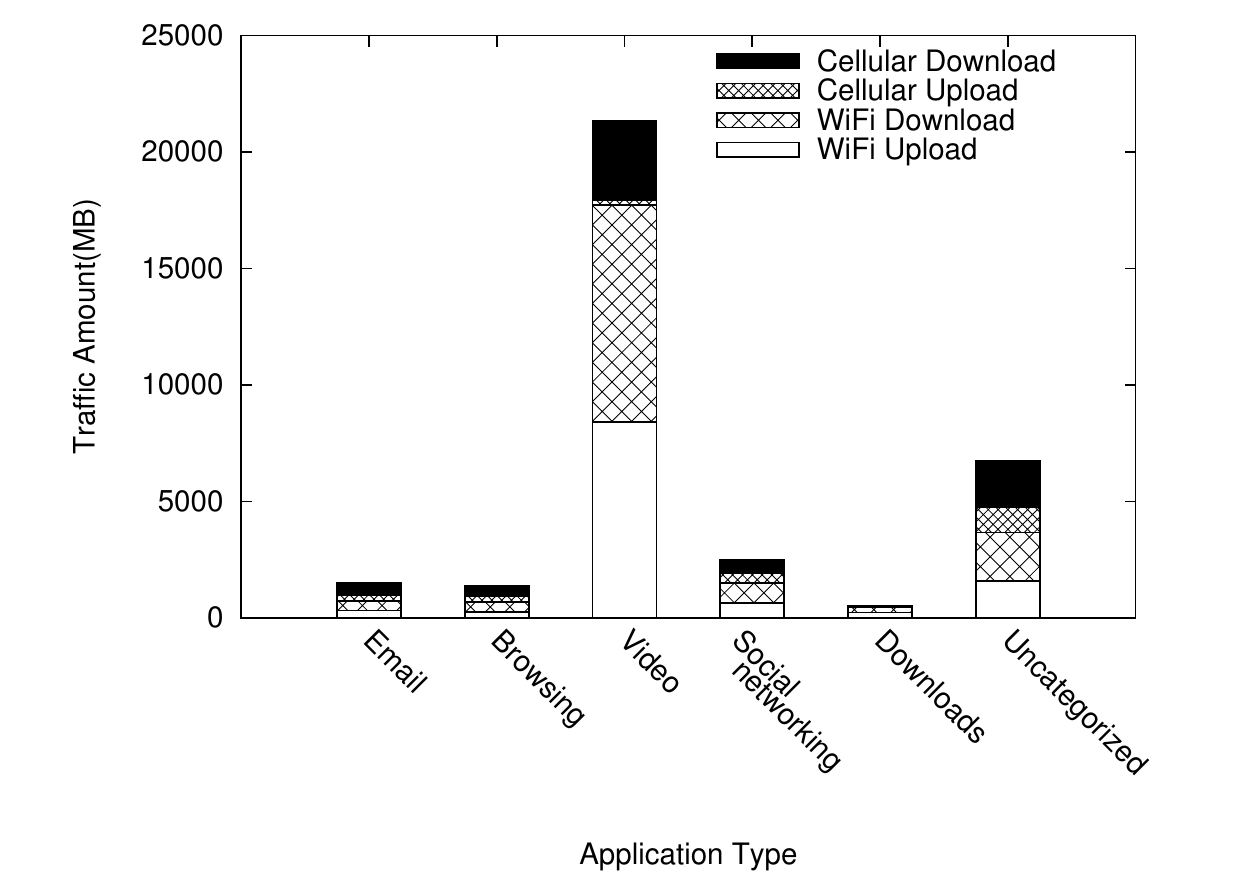}
\vspace{-0.1in}
\caption{Traffic amounts for each category and network type.}
\label{fig:category_traffic_portion}
\end{figure}

\begin{figure}[t]
\vspace{-0.2in}
\centering
\includegraphics[scale=0.45]{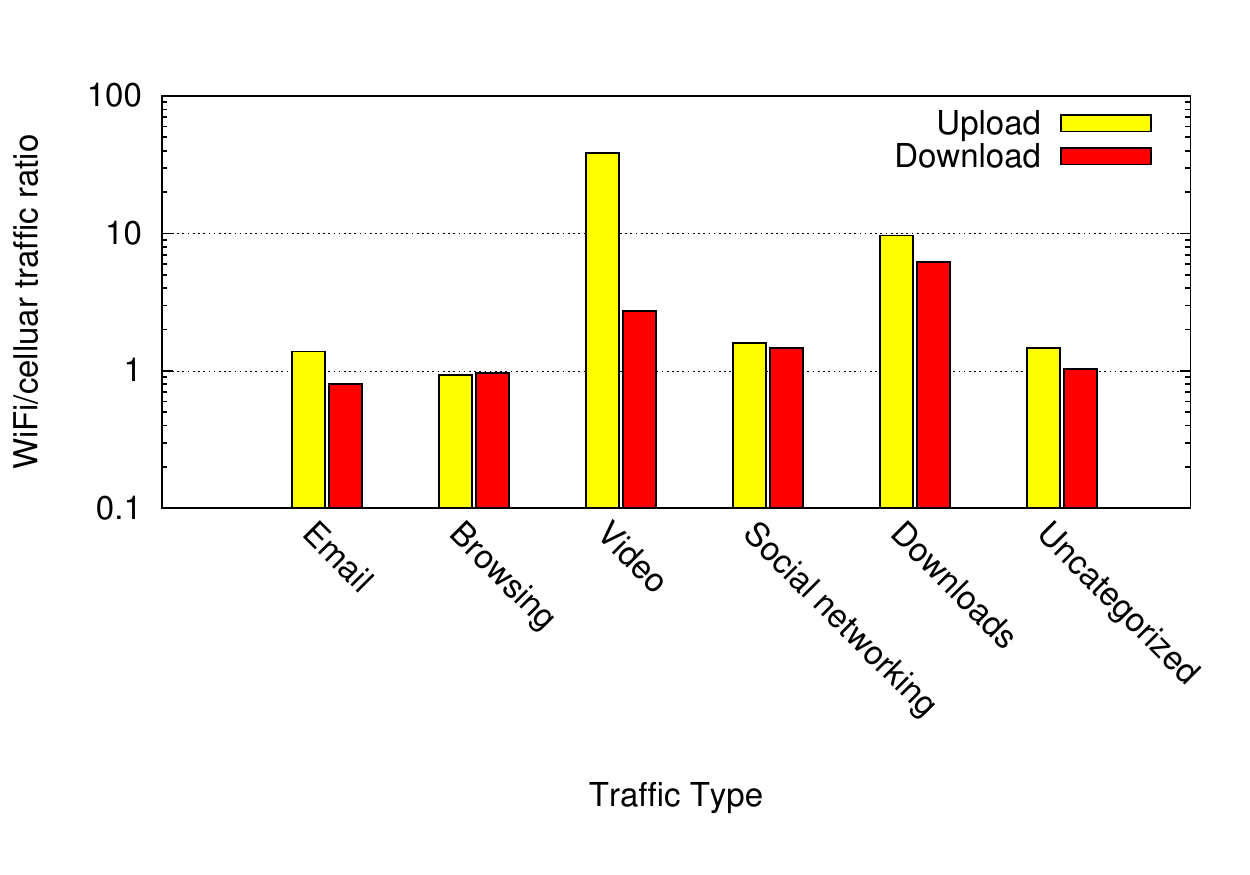}
\vspace{-0.1in}
\caption{The ratio of traffic WiFi to cellular network traffic for each application type.}
\label{fig:wifi_cellular_ratio}
\end{figure}
From users' usage traces, we find that users are already offloading a large portion of their traffic, but a large amount of video and social networking cellular traffic still runs over WiFi and can be delayed for offloading.  
In Figure \ref{fig:category_traffic_portion}, we illustrate the amount of upload and download traffic for each application type in cellular and WiFi networks. As expected, the amount of WiFi traffic is much larger than that of cellular, and the amount of download traffic is larger than the upload traffic. 

As in \cite{CiscoVNI2014}, video traffic accounts for the largest portion of traffic for WiFi uploads/downloads and cellular downloads (74, 70, 49\% respectively). For these types of traffic, the order of the application types according to the traffic amount is Video, Social networking, Email, Browsing, and Downloads. In the case of cellular upload, the traffic amount is in the order of Social networking-Browsing-Email-Video-Downloads. We find that 84\% of the total upload and 66\% of the total download traffic uses WiFi.

To investigate the fraction of each application's traffic that is offloaded to WiFi networks, we calculate the ratio of WiFi to cellular upload and download traffic for all the application types, as shown in Figure \ref{fig:wifi_cellular_ratio}. If this ratio is large, it means that the users already offload their traffic to WiFi for that application type, either because that application type is delay-tolerant or because it requires WiFi's high bandwidth (e.g., video applications). 

From Figure \ref{fig:wifi_cellular_ratio}, we see that different application types show different ratios. In particular, the Video and Download types show large values for both upload and download traffic. This coincides with the small values of $\mu$ in Table \ref{tab:param} for these application types. Video requires high bandwidth and has high cost generally, incentivizing users to wait for WiFi in order to use high bandwidth. However, by comparing the ratios for Video and Downloads, we observe that users wait more for Video than for Downloads. While a significant fraction of video and downloads are offloaded, the high delay tolerance of Download apps indicates that more offloading is possible.

%

%

\begin{figure}[t]
\vspace{-0.2in}
\centering
\includegraphics[scale=0.45]{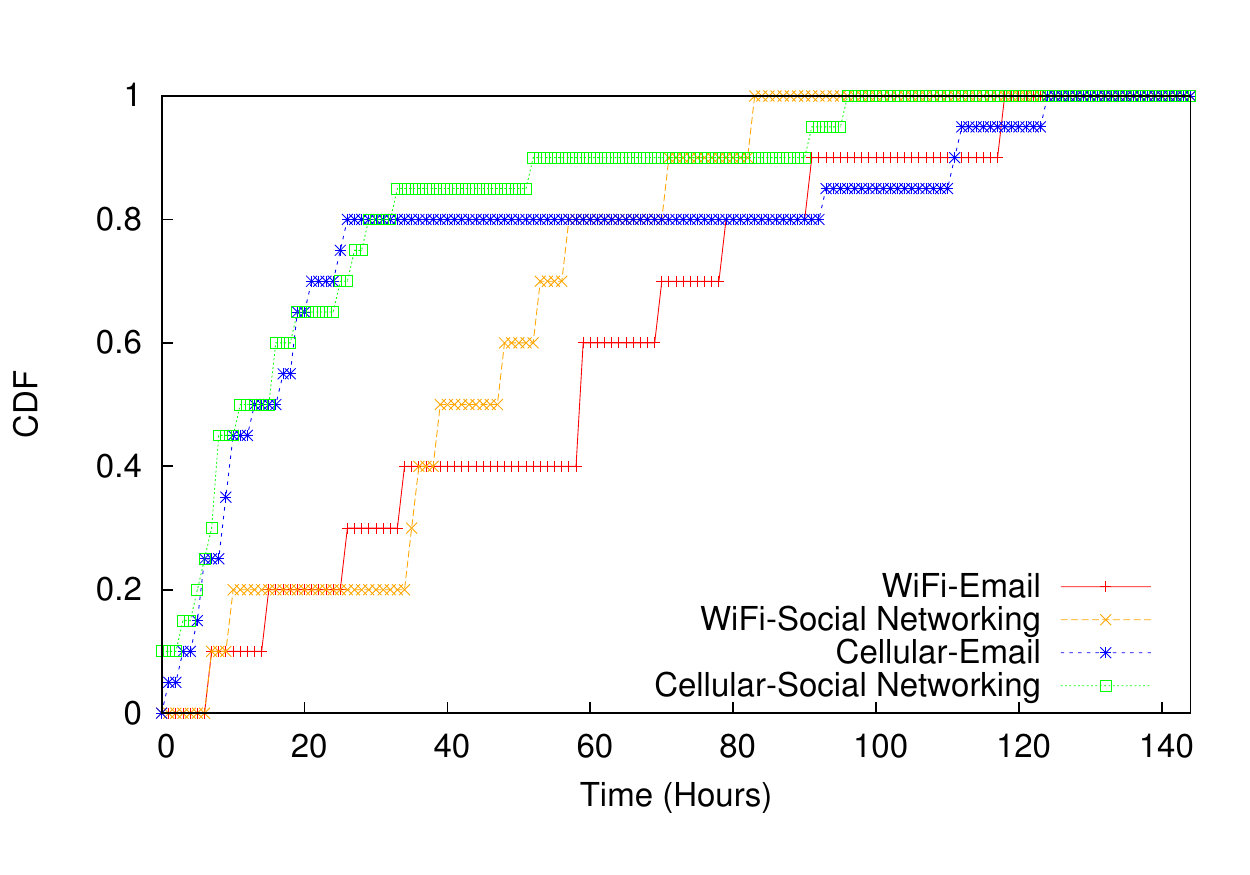}
\vspace{-0.1in}
\caption{CDF of the number of periods for which each user uses Email and Social networking applications.}
\label{fig:application_usage_cdf}
\end{figure}


We also see that the Email and Social networking applications have some offloading potential. 
Figure \ref{fig:application_usage_cdf} shows CDFs for the number of periods in which a user uses E-mail and Social networking applications. We observe that about 40\% of users use WiFi data for Email and Social networking applications 30\% of the time. Over cellular, the data usage frequency decreases, but 20\% of users spend significant amounts of time on email and social networking. This high usage frequency indicates that some Email and Social Networking traffic can be offloaded if a user does not need to wait very long for WiFi access. 

\section{Experimental Evaluation} \label{sec:evaluation}

To evaluate the effects of AMUSE's Bandwidth Optimizer (Algorithm \ref{alg:everything} in Section \ref{sec:allocation_algorithm}) on users' offloading experience, we collected 3G and WiFi usage and mobility data from an additional 12 Android and 25 iPhone users over a period of 19 days and one week, respectively, as explained in Section \ref{sec:data_collection}.  We then simulate the performance of AMUSE's bandwidth optimizer, taking the recorded usage data as the historical usage, and compare AMUSE's performance with two other known offloading algorithms \revise{in \cite{OffloadingCoNext2010}}.  Our results show that AMUSE can both reduce users' spending and improve their utility compared with these two benchmarks.

\subsection{Experimental Data and Settings} \label{sec:empirical_data_collection}

\begin{figure*}[ht]
\vspace{-0.2in}
\centering
\subfloat[Cellular usage.]{\includegraphics[scale=0.45,bb = 30 40 360 202]{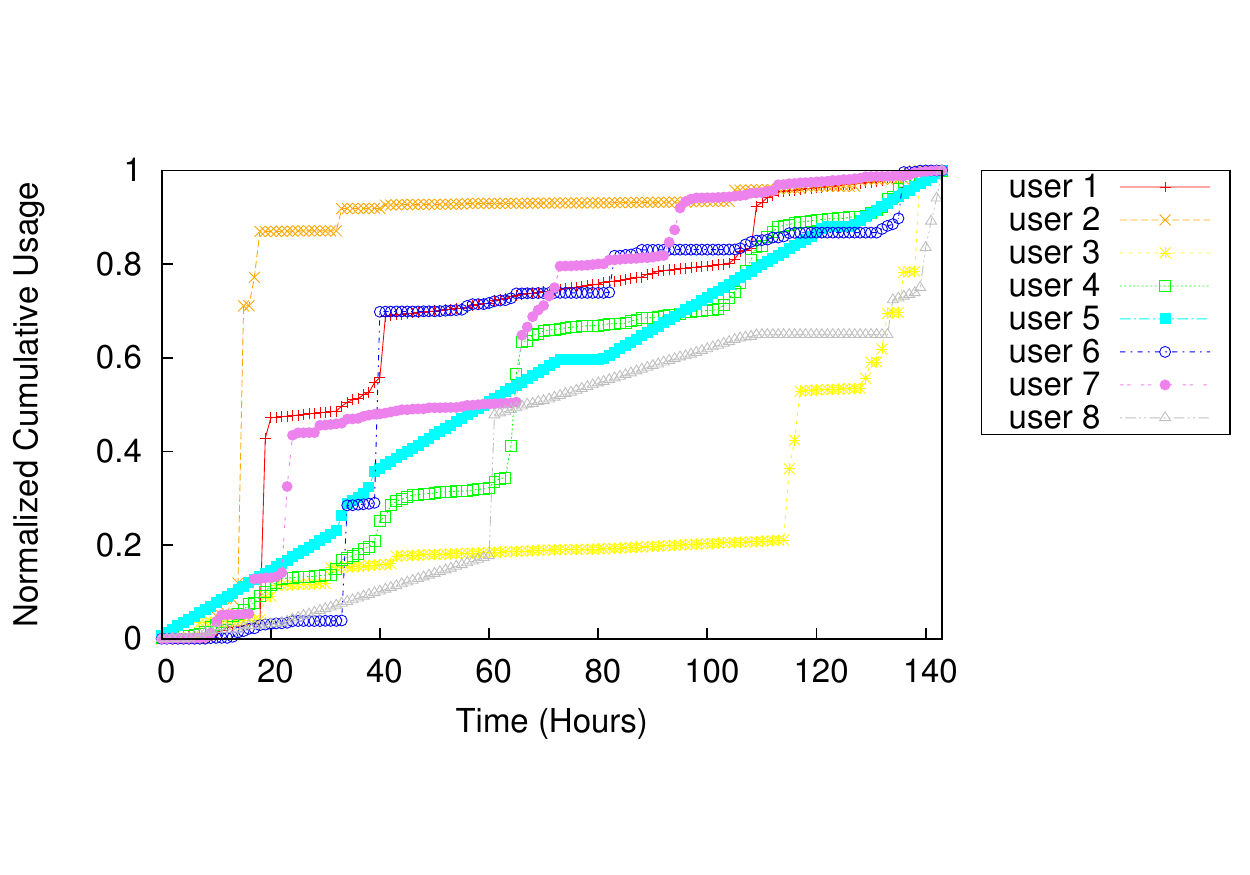}\label{fig:cumulative_usage_iPhone_cellular}}
\subfloat[WiFi usage.]{\includegraphics[scale=0.45,bb = -30 40 320 202]{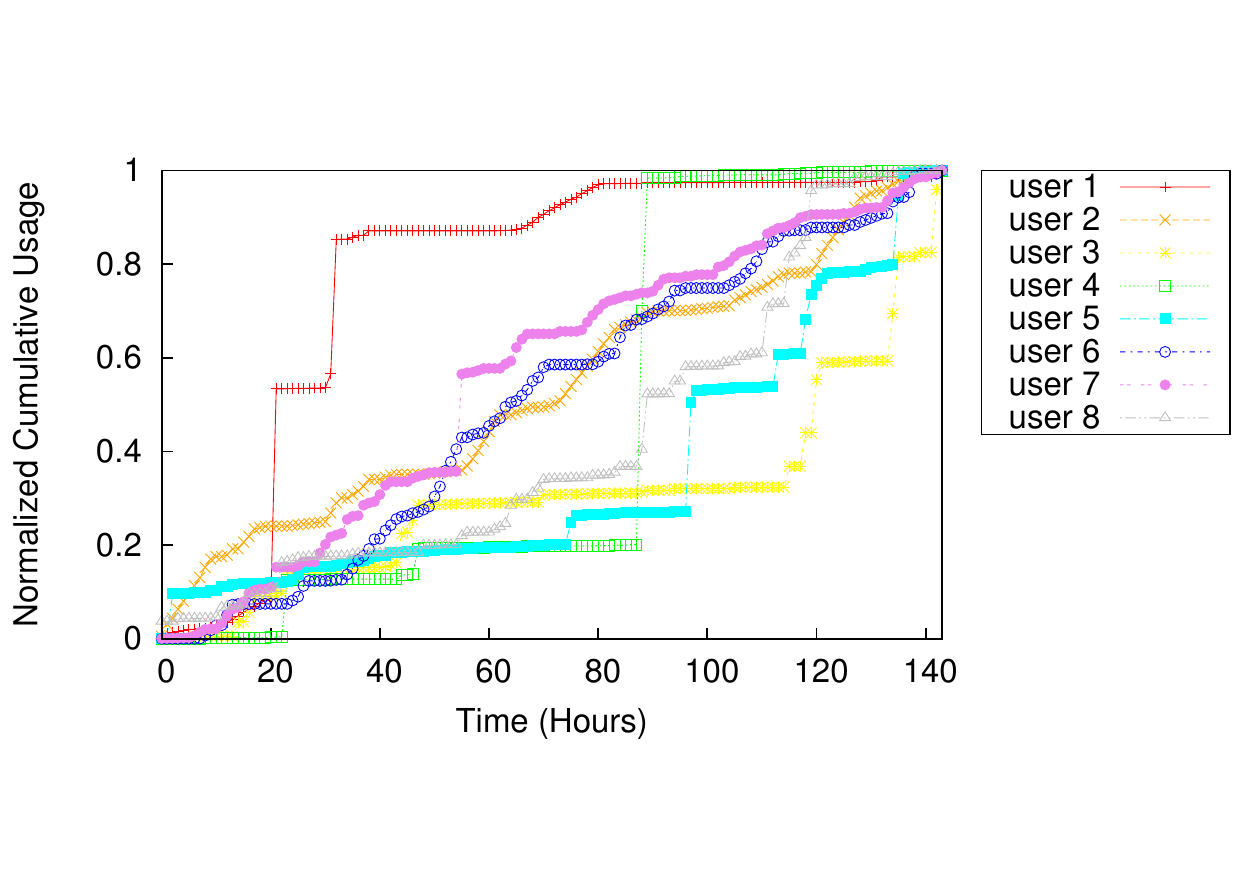}\label{fig:cumulative_usage_iPhone_wifi}} 
\caption{Normalized cumulative usage of iPhone users.}
\end{figure*}

\begin{figure*}[ht]
\vspace{-0.2in}
\centering
\subfloat[Cellular usage.]{\includegraphics[scale=0.45,bb = 30 40 360 202]{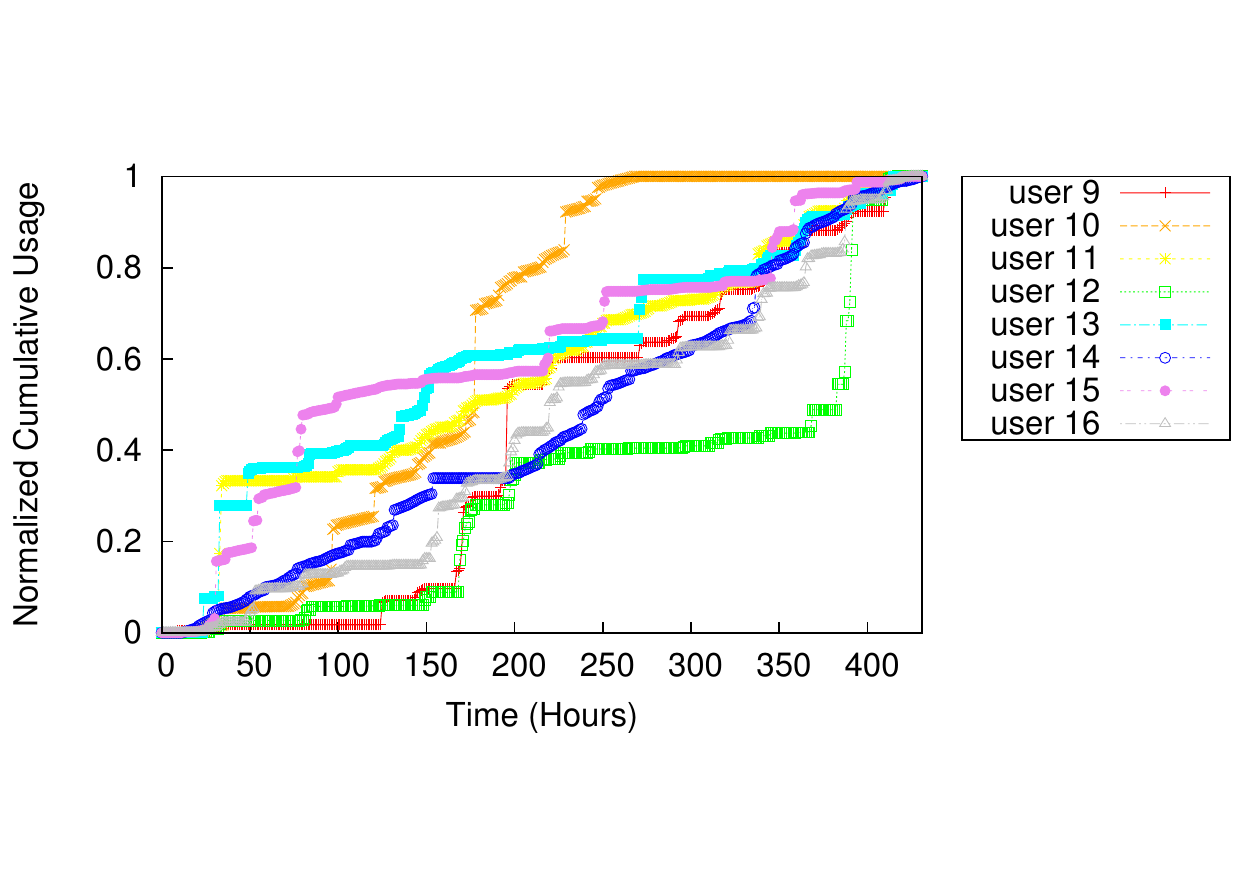}\label{fig:cumulative_usage_android_cellular}}
\subfloat[WiFi usage.]{\includegraphics[scale=0.45,bb = -30 40 320 202]{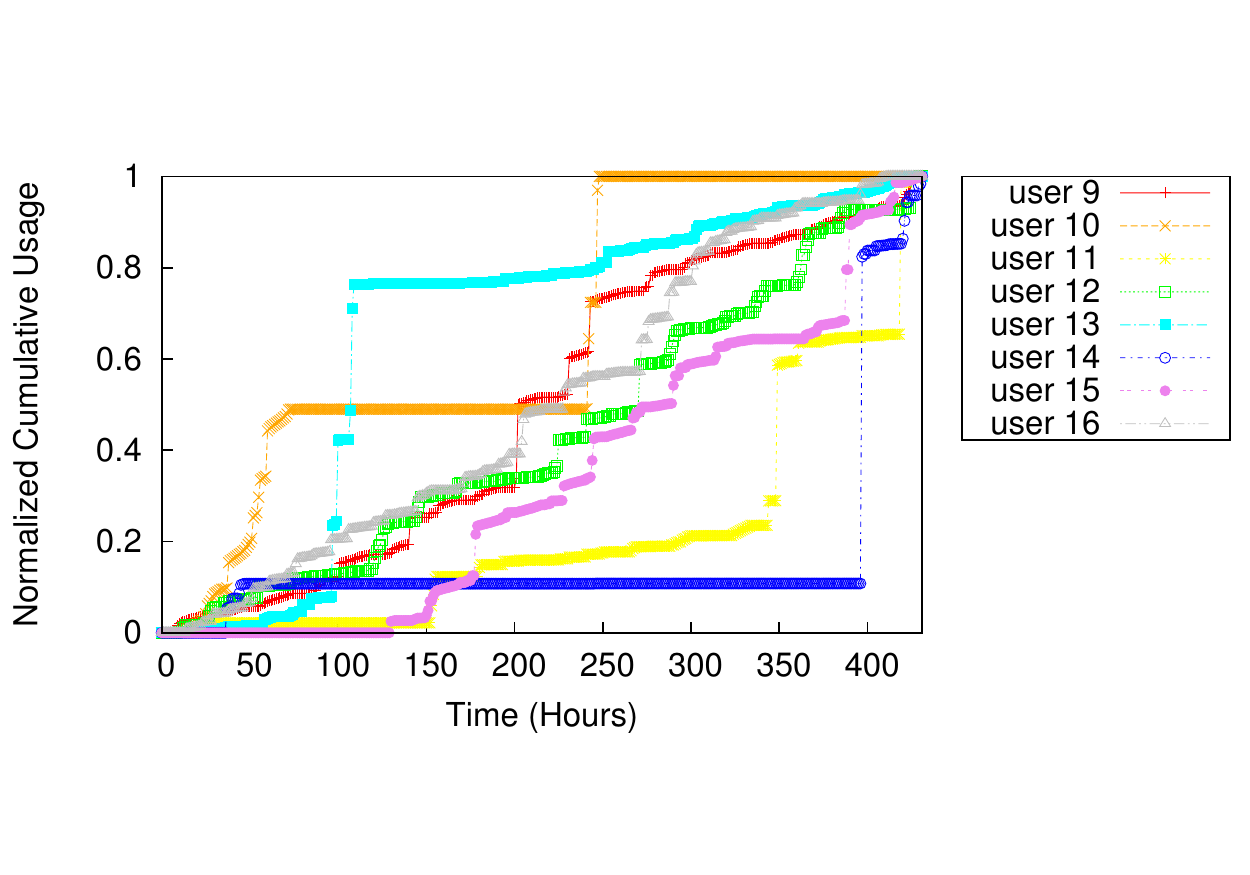}\label{fig:cumulative_usage_android_wifi}} 
\caption{Normalized cumulative usage of Android users.}
\end{figure*}

%

%

Since some of our users exhibited very similar traffic patterns and some showed very limited data usage, we choose sixteen representative users' data on which to run the AMUSE simulation (eight each for iPhone and Android). 
Figures \ref{fig:cumulative_usage_iPhone_cellular} and \ref{fig:cumulative_usage_iPhone_wifi} represent the normalized cumulative usage of selected iPhone users for cellular and WiFi, respectively. 
The normalized cumulative usage of selected Android users for cellular and WiFi networks are shown in Figure \ref{fig:cumulative_usage_android_cellular} and \ref{fig:cumulative_usage_android_wifi}. 
We can observe that the chosen representative users show diverse usage patterns in both WiFi and cellular networks.
We use three days of data as each user's usage history, and run the simulation assuming hour-long periods.  The user's monthly budget for 3G data usage is chosen from a truncated normal distribution between \$20  and \$40 (2 to 4 GB at a unit price of \$10/GB).


\begin{figure}[t]
\vspace{-0.2in}
\centering
\includegraphics[scale=0.4]{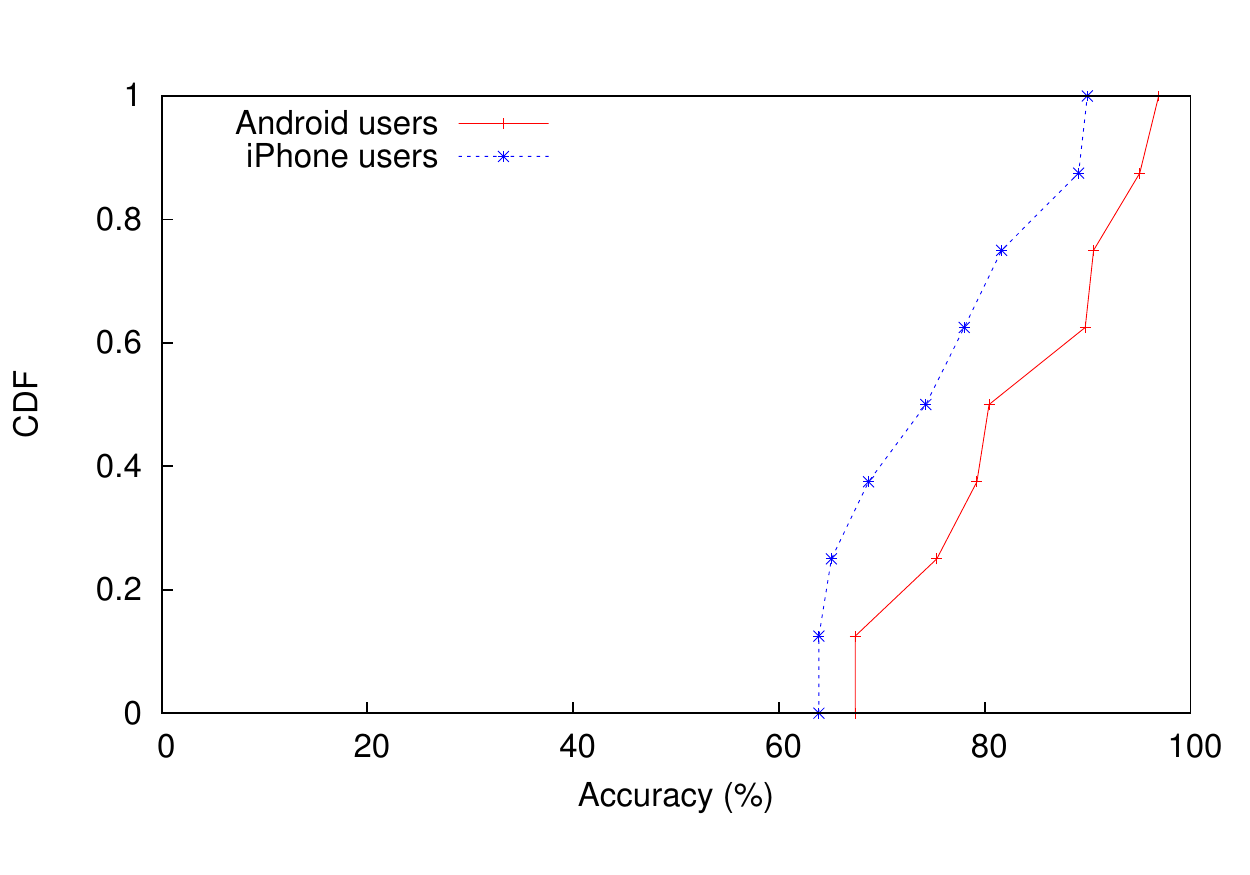}
\vspace{-0.2in}
\caption{CDF of WiFi prediction accuracy.}
\label{fig:wifi_prediction_accuracy}
\end{figure}
\begin{figure}[t]
\vspace{-0.2in}
	\centering
	\includegraphics[scale=0.4]{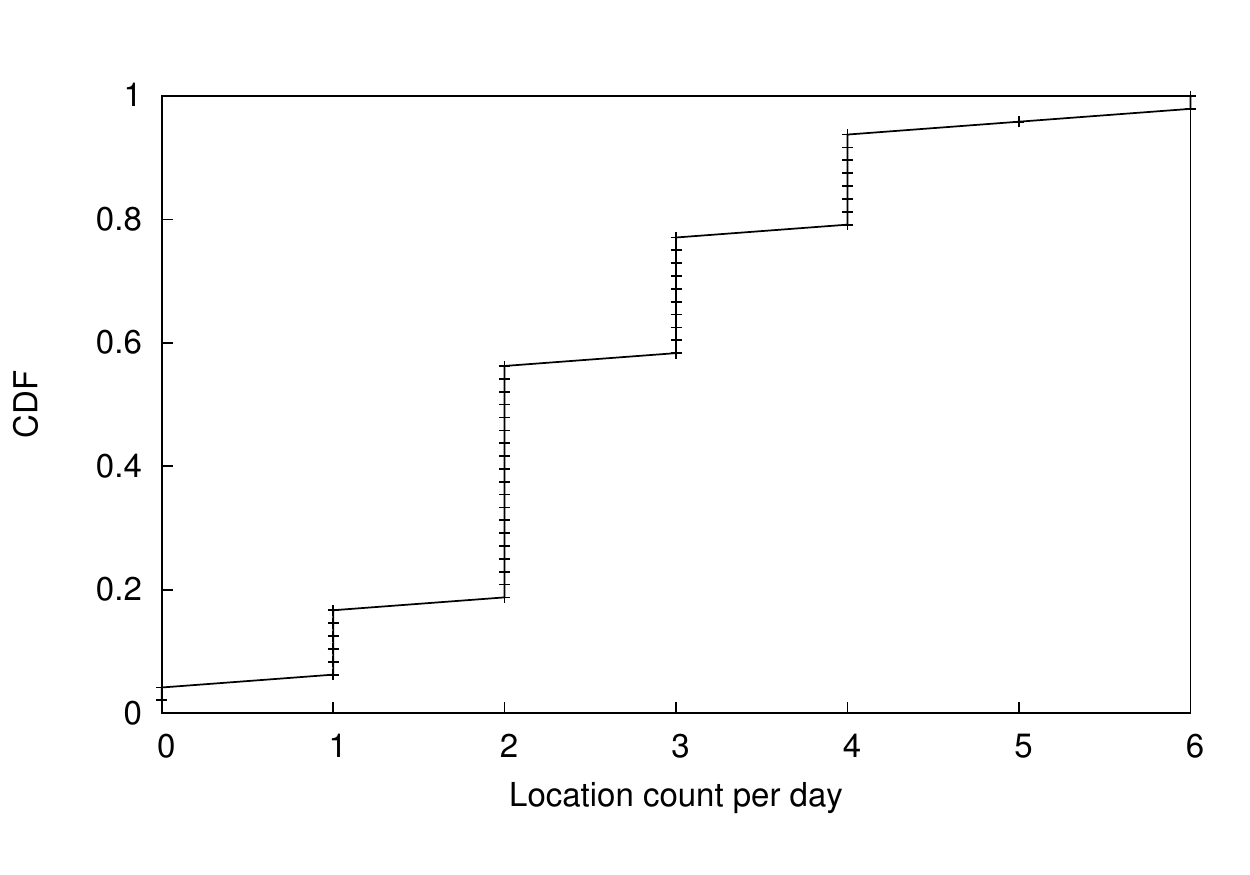}
	\vspace{-0.2in}
	\caption{CDF of the number of different locations per a day where WiFi network is accessed.}
	\label{fig:cdf_wifi_location_count}
\end{figure}
\begin{figure}[t]
	\centering
	\includegraphics[scale=0.4,angle=-90,bb=79 31 310 599]{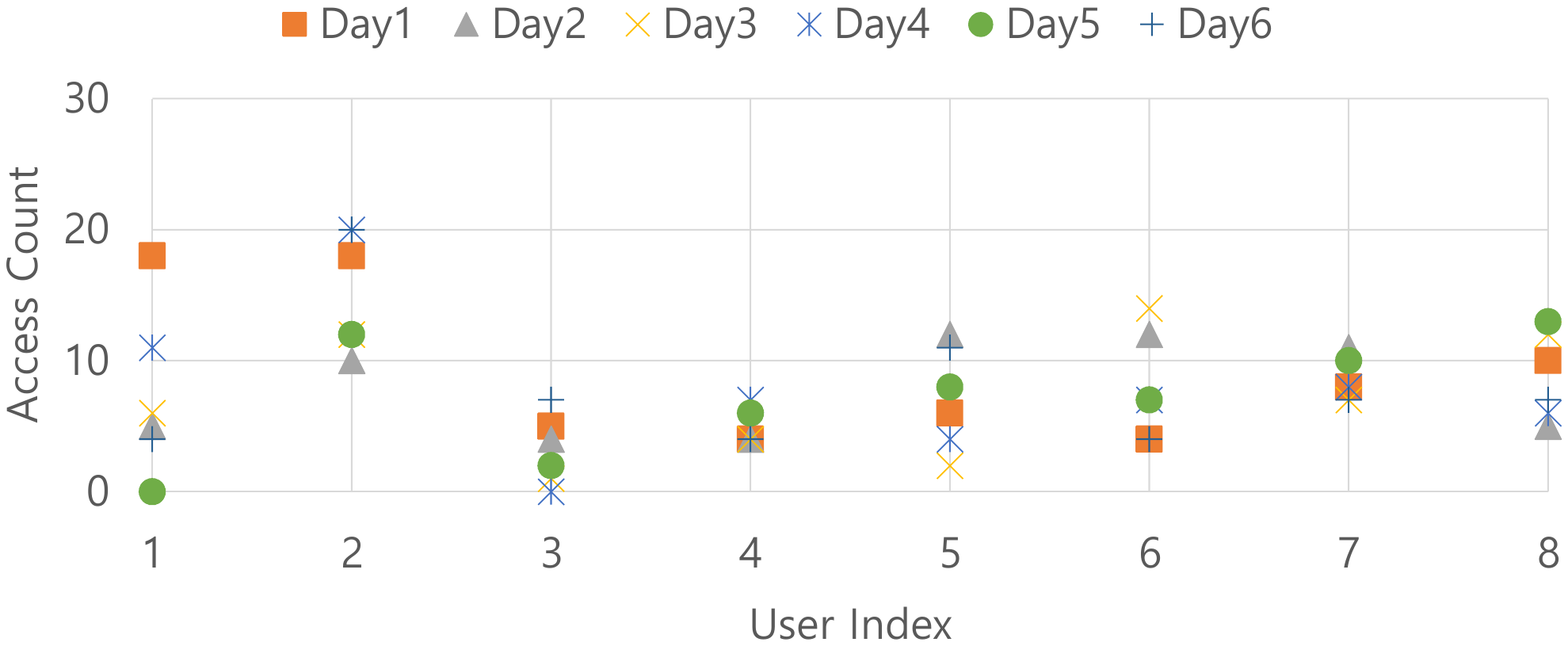}
	\caption{WiFi access frequency per a day of each iPhone user.}
	\label{fig:wifi_access_count}
\end{figure}

To verify that our three-day training set of data is \revise{reasonable} to simulate AMUSE, we test the accuracy of our WiFi prediction algorithm on this dataset.  To do so, we define the prediction accuracy as follows: if the probability of WiFi access for a given user in a given future period is greater (respectively less) than 0.5 and we observe (respectively do not observe) WiFi access in that period, we classify the prediction as ``accurate.''  Otherwise, we call the prediction ``inaccurate.''  We then divide the number of accurate predictions by the total number of predictions for each user to find the prediction accuracies. \revise{As shown in Figure \ref{fig:wifi_prediction_accuracy}, we observe a 64 -- 90\% prediction accuracy for the eight iPhone users and a 67 -- 97\% accuracy for the eight Android users. 
To further inspect the prediction accuracy, we tested the algorithm on other open mobility traces. We used the data of 21 users who have enough records needed for prediction from the dataset collected in the LifeMap project~\cite{chon2011mobility}. We obtained similar accuracy as on our dataset (64 -- 95\% accuracy).}

\revise{We show the WiFi access frequency per a day of each iPhone user in Figure~\ref{fig:wifi_access_count}. Except user 1, each user has a similar WiFi access frequency for each day. 
We show the number of different locations per a day where WiFi network is accessed in Figure~\ref{fig:cdf_wifi_location_count}. We can observe that each user accesses the WiFi network within a small number of locations. These results coincides with assumptions of our location dependent WiFi prediction method.}  



\subsection{Baseline Algorithms}

We compare AMUSE's performance to two baseline algorithms: on-the-spot offloading and delayed offloading \cite{OffloadingCoNext2010}. 

On-the-spot offloading offloads traffic to WiFi opportunistically: users send their traffic over WiFi if they are connected to a WiFi network at that time, and switch to 3G if they move outside of the WiFi coverage area.  No sessions ever wait for WiFi, which may lead to higher spending as compared with AMUSE -- AMUSE allows delay-tolerant sessions to wait for WiFi, thus saving users money.
The delayed offloading algorithm forces all traffic to wait up to a fixed time limit for WiFi access (1 hour in our simulations). If WiFi becomes available before this time, the waiting traffic is sent over WiFi; otherwise, it is sent over 3G. While this algorithm may offload more traffic than AMUSE and thus save users money, it does not consider users' delay tolerances: even urgent sessions are forced to wait for some time.  Moreover, if WiFi is not available at the end of the fixed time limit, the sessions will complete over 3G anyway, costing users money and making them wait.  


%

\begin{figure}[t]
\vspace{-0.1in}
\centering
\includegraphics[scale=0.55]{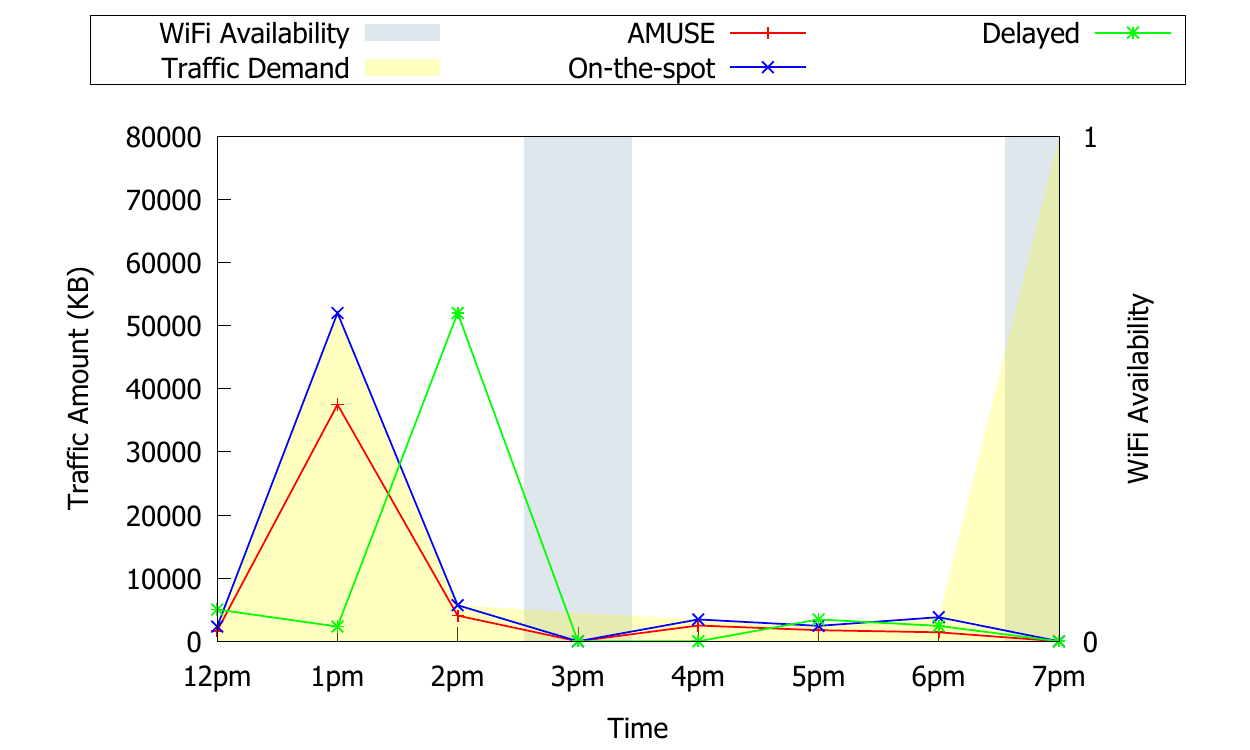}
\vspace{-0.1in}
\caption{An example traffic demand and cellular usage amount under each offloading algorithm.}
\label{fig:traffic_demand_example_combined}
\end{figure}

In Figure \ref{fig:traffic_demand_example_combined}, we show an illustrative example scenario that explains how AMUSE operates compared to other algorithms in making offloading decisions. Figure \ref{fig:traffic_demand_example_combined} illustrates the cellular traffic demand and WiFi availability of one user for 8 hours. At 1pm, the traffic demand is larger than other times, and WiFi is available at 3pm and 7pm. The amount offloaded with each offloading algorithm can be calculated from the difference in total demand and traffic for each algorithm. Since AMUSE predicts the WiFi availability and delays longer than 1 hour if the utility is increased, it offloads a considerable amount of the traffic in 1pm to 3pm. On the other hand, On-the-spot offloading offloads only the traffic at 3pm and 7pm when the WiFi is available. Delayed offloading offloads more than On-the-spot offloading by delaying the traffic of 2pm and 6pm to 3pm and 7pm, respectively. However, since it cannot predict the WiFi availability of 3pm at 1pm, it cannot delay the traffic of 1pm to 3pm, thus losing the opportunity to offload. Moreover, since it blindly delays the traffic by one hour if the WiFi is not available, the traffic of 12, 1, 4, 5pm is delayed but ultimately transmitted by 3G, decreasing users' utility. AMUSE, on the other hand, does not always prefer offloading to transmitting on 3G network. For example, AMUSE delays only some traffic at 1pm, since the expected utility of transmitting over 3G is larger than that of waiting for higher bandwidth and lower cost WiFi. 



\subsection{Numerical Results}

\begin{figure}[t]
\vspace{-0.3in}
\centering
\includegraphics[scale=0.45]{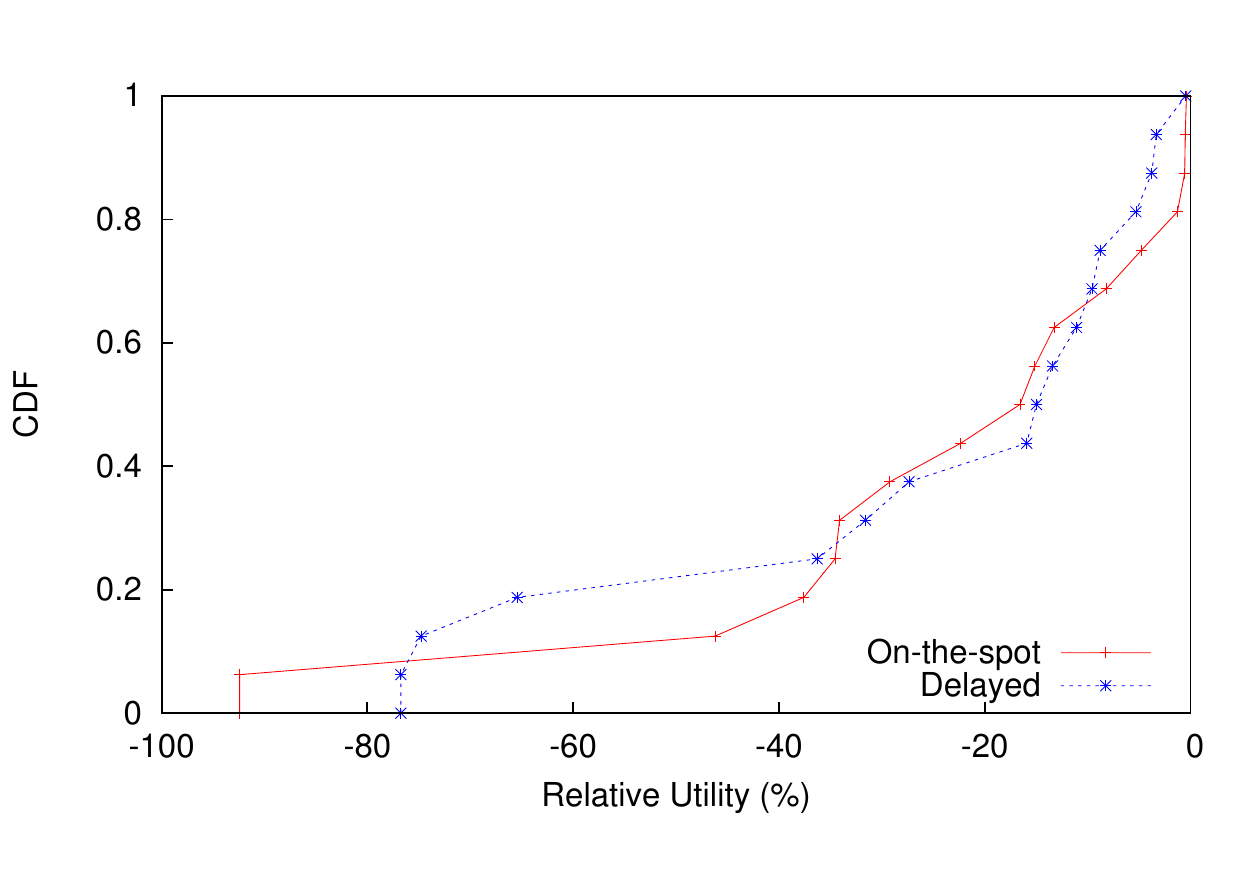}
\vspace{-0.1in}
\caption{CDF of relative utility function values compared to AMUSE.}
\label{fig:utilities}
\end{figure}

Figure \ref{fig:utilities} plots the distributions of relative utility values under our benchmark algorithms compared to those under AMUSE.  For each user, both benchmark algorithms decrease the utility. This decrease is particularly dramatic for one user, whose utility values with on-the-spot and delayed offloading are only 8 and 23\%, respectively, of the utility under AMUSE.  On average, the utility of on-the-spot offloading is 19\% less than that of AMUSE, while that of delayed offloading is 22\% lower.

\begin{figure}[t]
\vspace{-0.3in}
\centering
\includegraphics[scale=0.45]{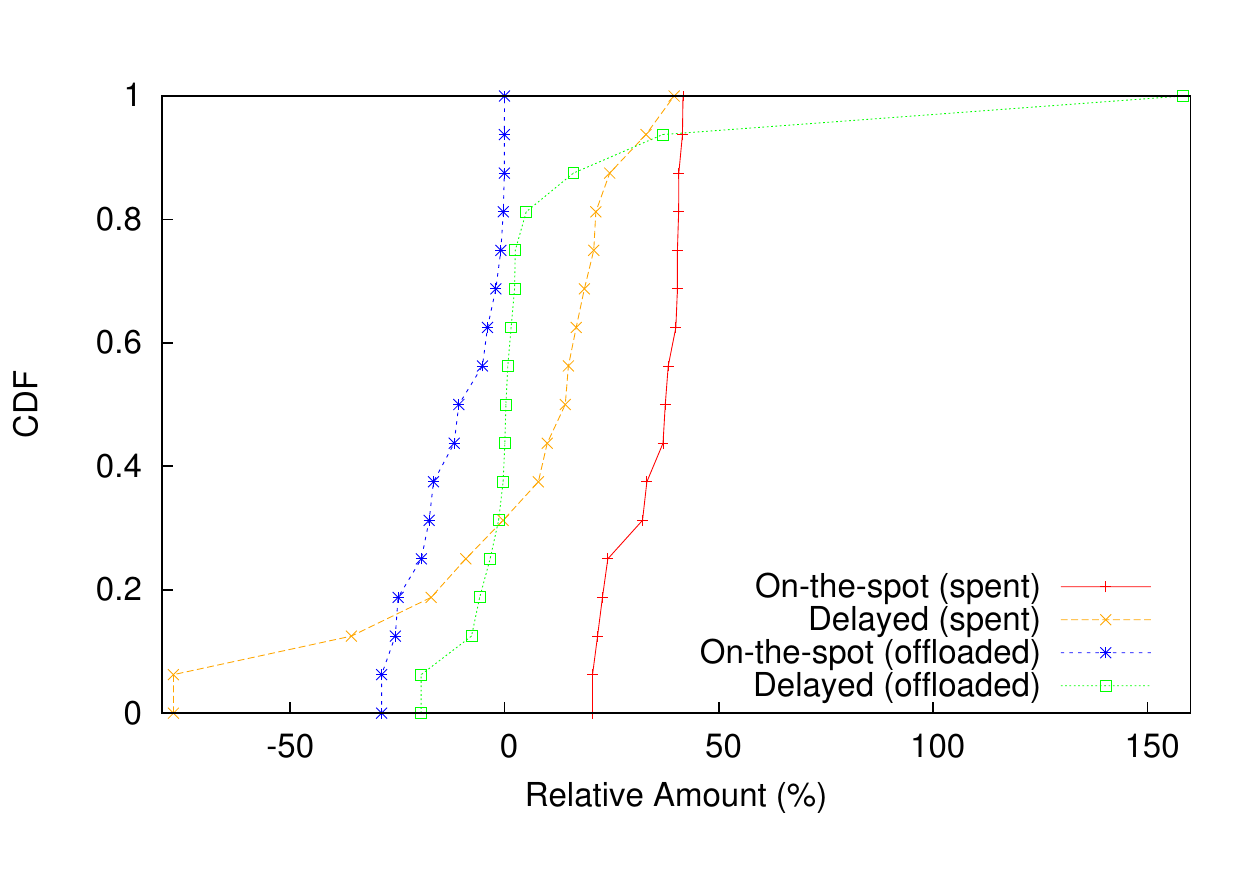}
\vspace{-0.1in}
\caption{CDF of relative offloaded traffic amount and amount spent compared to AMUSE.}
\label{fig:offloaded_spent_cdf}
\end{figure}

%

AMUSE yields higher utility values than on-the-spot offloading due to offloading more traffic onto WiFi: Fig. \ref{fig:offloaded_spent_cdf} shows the distributions of relative amounts of traffic offloaded under both benchmark algorithms, as compared to AMUSE.  We see that for all users, the amount of traffic offloaded is larger under AMUSE than it is under on-the-spot offloading; thus, AMUSE leverages the delay tolerance of some sessions by allowing them to wait for WiFi access.  Users then save money: Fig. \ref{fig:offloaded_spent_cdf} also compares users' amount spent under the two benchmark algorithms to that spent with AMUSE.  Users consistently spend over 20\% more with on-the-spot offloading, and on average increase their spending by 33\% compared with AMUSE.


Compared to delayed offloading, AMUSE trades off between reducing users' spending by offloading traffic and completing some sessions immediately due to their intolerance of delay.  Figure \ref{fig:offloaded_spent_cdf} shows that delayed offloading offloads more traffic than AMUSE for 10 users: AMUSE sends some sessions over 3G without waiting for WiFi, allowing users to spend more and delay less.  
One user offloads nearly 160\% more traffic under delayed offloading relative to AMUSE. The consequent decrease in cost relative to AMUSE (nearly 80\%) is offset by less delay under AMUSE; this user in fact experiences 5\% less utility under delayed offloading than that under AMUSE.
On the other hand, delayed offloading offloads less traffic than AMUSE for 6 users: AMUSE allows delay-tolerant traffic to wait more than an hour for WiFi. We found that this additional wait for WiFi reduces these users' spending, offsetting the loss in utility from delaying the session.

\begin{figure}[t]
\vspace{-0.1in}
\centering
\includegraphics[scale=0.45]{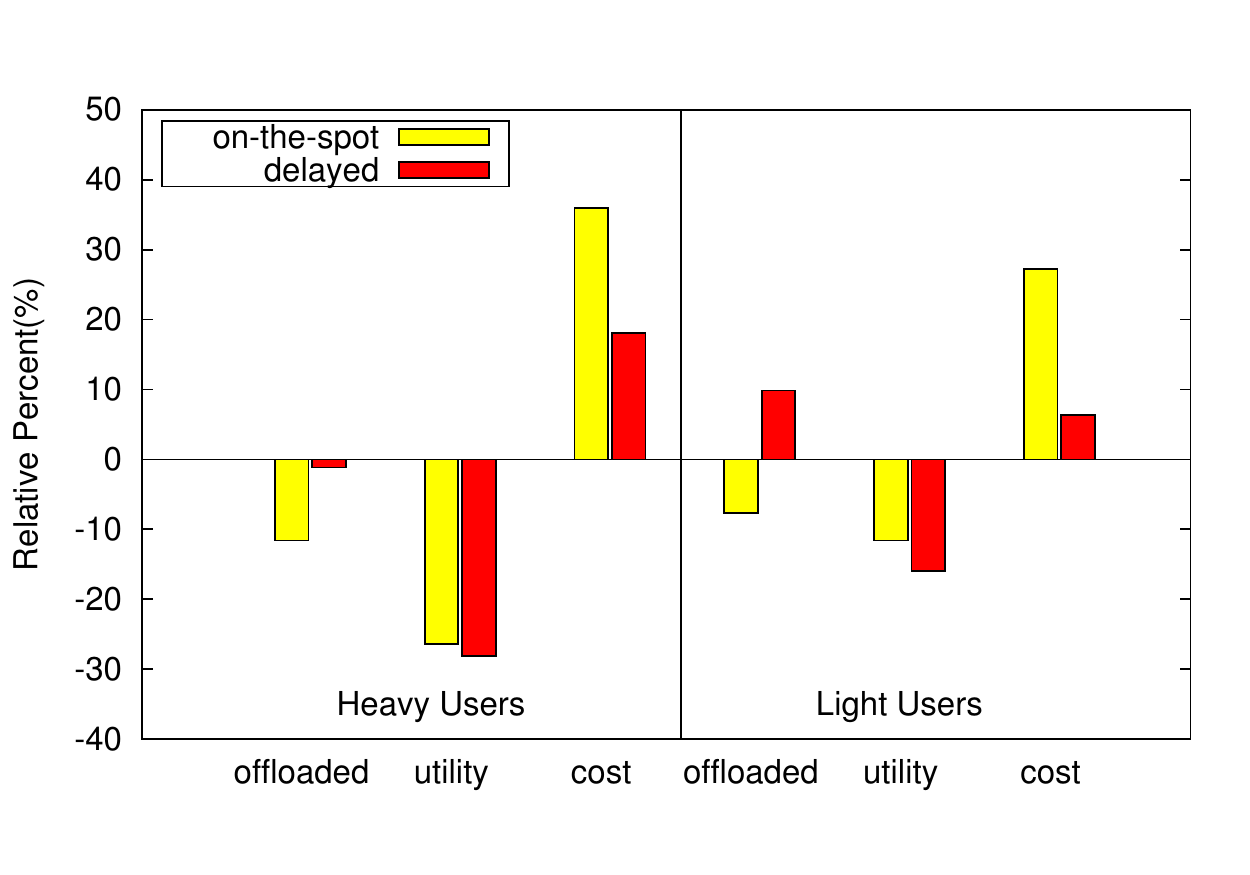}
\vspace{-0.25in}
\caption{Average offloaded traffic, utility, and cost of heavy and light users in all traces.}
\label{fig:total_iphone_android}
\vspace{-0.07in}
\end{figure}

Finally, we examine AMUSE's benefits for different types of users.  We split 16 users into ``heavy'' and ``light'' usage groups (8 users for each group), and plot the amount offloaded, utility, and cost of the two benchmark algorithms relative to AMUSE in Fig. \ref{fig:total_iphone_android}.  For heavy users, both benchmark algorithms perform worse than AMUSE: users' utility and amount offloaded decrease under these algorithms, while their cost increases.  Thus, AMUSE's cost savings in offloading more traffic offset any loss in utility from waiting more for WiFi access.  On-the-spot offloading performs especially poorly compared to AMUSE, indicating that most users' delay tolerance enables them to gain utility under AMUSE by selectively delaying some sessions and sending them over WiFi. For light users, AMUSE weights the cost savings from waiting for WiFi less heavily: some sessions do wait for WiFi, as shown by the decrease in amount offloaded in on-the-spot offloading, but delay-intolerant sessions do not wait, as shown by the increase in amount offloaded in delayed offloading.  This likely arises from light users' looser budget constraint: by definition, light users spend less than heavy users on their data consumption.  They accordingly benefit less overall: compared with AMUSE, the utility of heavy users decreases by 27\% under the benchmark algorithms, while that of light users decreases by 14\%.

\section{Conclusion}
\label{sec:Conclusion}

In this paper, we propose AMUSE, a cost-aware WiFi offloading system that maximizes the {\it end user's} utility under her 3G budget constraints. AMUSE consists of two main components: a bandwidth optimizer and a TCP rate controller.  By predicting future usage and WiFi availability, the bandwidth optimizer chooses how long an application should wait for WiFi access, as well as a 3G data rate should WiFi not be available. These choices are optimized so as to balance the user's tradeoffs between the cost of sending an application's traffic over 3G, the higher throughput received over WiFi, and the delay inherent in waiting for WiFi. The TCP rate controller practically enforces the 3G rates chosen for each application by controlling the TCP advertisement window from the user side.  AMUSE also allows for end-user interaction by providing a user interface through which users can set their bandwidth allocation preferences and view the offloading decisions made.
Through a measurement study, we show that though a large amount of some applications' traffic is offloaded already, our offloading framework can offload a larger portion of mobile users' cellular traffic.

We prototyped AMUSE and evaluated its performance with mobile traces from 37 users. Our results show that AMUSE can improve both heavy and light data users' utility from offloading; for heavy users, two other representative WiFi offloading algorithms achieve 27\% lower utility than AMUSE on average.  Heavy users' costs were on average 18 and 36\% higher under these benchmark algorithms compared to AMUSE, a savings realized by offloading more traffic onto WiFi. Though our results are based on data from a limited number of users, we expect similar performance from a wider range of users. 


%



\ifCLASSOPTIONcompsoc
  \section*{Acknowledgments}
\else
  \section*{Acknowledgment}
\fi
{\footnotesize
This research was supported by Basic Science Research Program through the "National Research Foundation of Korea(NRF)" funded by the Ministry of Science, ICT \& future Planning (2013R1A2A2A01016562). 
This work was in part supported by NSF grants CNS 1011962 and CNS 1456847.}


\ifCLASSOPTIONcaptionsoff
  \newpage
\fi



\bibliographystyle{IEEEtran}
\raggedright
\bibliography{0_main_file.bbl}
\justifying
\vspace{-1.2in}
\begin{biography}[{\includegraphics[width=1in,height=1.25in,clip,keepaspectratio]{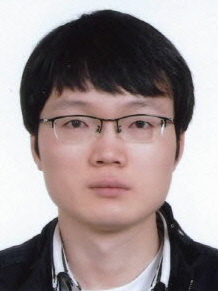}}]{Youngbin Im}
	is a postdoctoral researcher in the Department of Computer Science at University of Colorado Boulder. He received his B.S. and Ph.D. degrees in computer science and engineering from Seoul National University in 2006 and 2014, respectively. During his graduate program, he was a visiting student at Princeton University. His research interest includes mobile data offloading, next-generation Internet, multi-core based content router, video rate adaptation. 
\end{biography}	
\vspace{-0.7in}
\begin{biography}[{\includegraphics[width=1in,height=1.25in,clip,keepaspectratio]{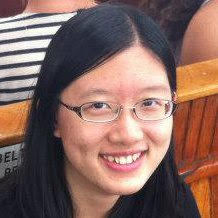}}]{Carlee Joe-Wong}
	is a Ph.D. candidate at Princeton University's Program in Applied and Computational Mathematics. Her research interests include network economics and optimal control. She received her A.B. in mathematics in 2011 and her M.A. in applied mathematics in 2013, both from Princeton University.  In 2013, she was the Director of Advanced Research at DataMi, a startup she co-founded in 2012 that commercializes new ways of charging for mobile data. She received the INFORMS ISS Design Science Award in 2014 and the Best Paper Award at IEEE INFOCOM 2012. In 2011, she received a National Defense Science and Engineering Graduate Fellowship.
\end{biography}	
\vspace{-0.7in}
\begin{biography}[{\includegraphics[width=1in,height=1.25in,clip,keepaspectratio]{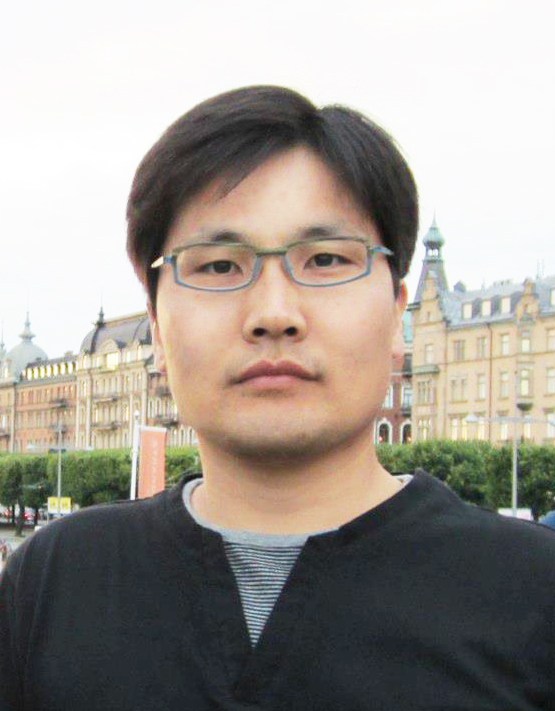}}]{Sangtae Ha} (S'07, M'09, SM'12) is an Assistant Professor in the
	Department of Computer Science at the University of Colorado at
	Boulder. He received his Ph.D. in Computer Science from North Carolina
	State University. His research focuses on building and deploying
	practical systems. He is a co-founder and the founding CTO/VP
	Engineering of DataMi, a startup company on mobile networks, and is a
	technical consultant to a few startups. He is an IEEE Senior Member
	and serves as an Associate Editor for IEEE Internet of Things (IoT)
	Journal. He received the INFORMS ISS Design Science Award in 2014.
\end{biography}	
\vspace{-0.7in}
\begin{biography}[{\includegraphics[width=1in,height=1.25in,clip,keepaspectratio]{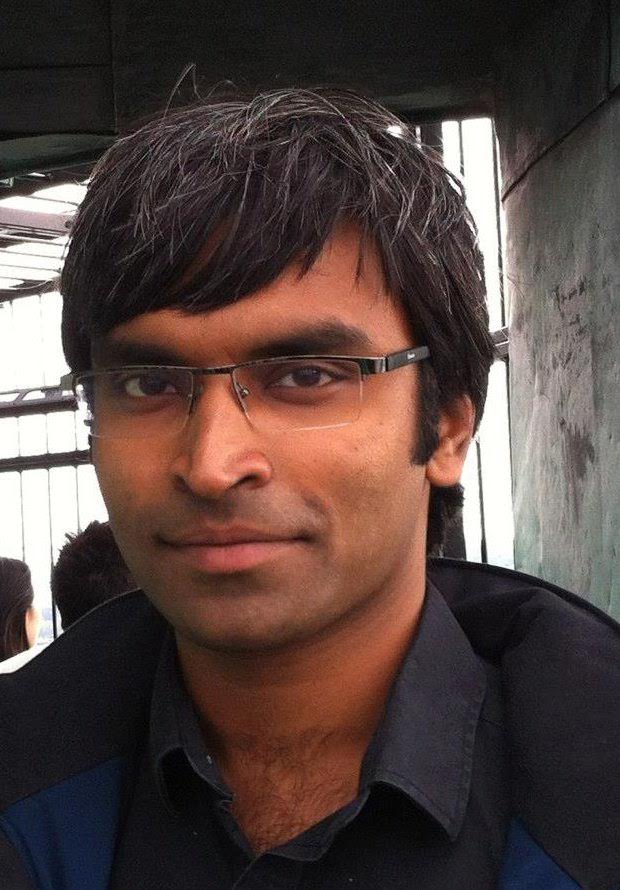}}]{Soumya Sen}
	received his M.S. and Ph.D. in Electrical and Systems Engineering from the University of Pennsylvania in 2008 and 2011, respectively, and was a postdoctoral researcher at the Princeton University. He is currently an Assistant Professor in the Department of Information \& Decision Sciences at the Carlson School of Management of the University of Minnesota. His research interests are in network technologies and e-commerce. He co-founded a telecom startup, DataMi, and served as the editor for the book, "Smart Data Pricing", published by John Wiley \& Sons in 2014. He won the Best Paper Award at IEEE INFOCOM 2012 and the INFORMS ISS Design Science Award 2014.
\end{biography}	
\vspace{-0.7in}
\begin{biography}[{\includegraphics[width=1in,height=1.25in,clip,keepaspectratio]{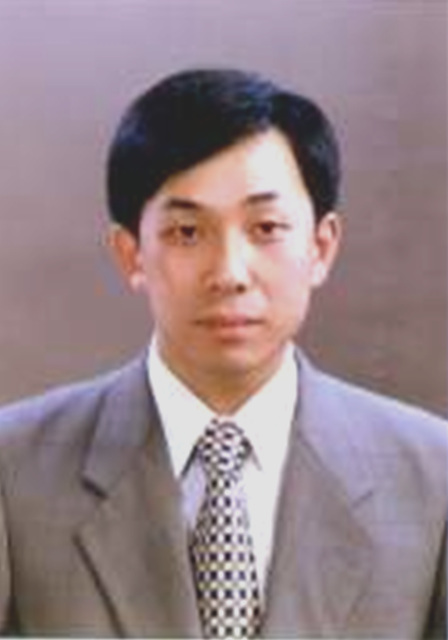}}]{Ted ``Taekyoung'' Kwon}
	is a professor with Department of Computer Science and Engineering, Seoul National University (SNU). 
	Before joining SNU, he was a Postdoctoral Research Associate at University of California Los Angeles and 
	City University New York. He obtained BS, MS and PhD at SNU in 1993, 1995, 2000, respectively. During his graduate program, 
	he was a visiting student at IBM T.J. Watson Research Center and at University of North Texas. He was a visiting professor 
	at Rutgers University in 2010.  His research interest lies in future Internet, indoor localizatino, network security, and wireless networks.
\end{biography}
\vspace{-0.7in}
\begin{biography}[{\includegraphics[width=1in,height=1.25in,clip,keepaspectratio]{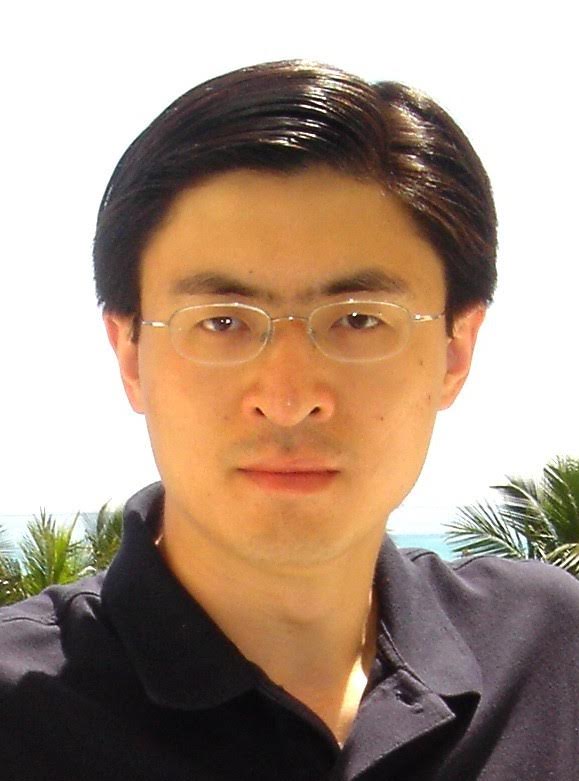}}]{Mung Chiang}
	(S'00, M'03, SM'08, F'12) is the Arthur LeGrand Doty Professor of Electrical Engineering at Princeton University and the recipient of the 2013 Alan T. Waterman Award. He created the Princeton EDGE Lab in 2009 to bridge the theory-practice divide in networking by spanning from proofs to prototypes, resulting in a few technology transfers to industry, several startup companies and the 2012 IEEE Kiyo Tomiyasu Award. He serves as the inaugural Chairman of Princeton Entrepreneurship Council and the Director of Keller Center for Innovation in Engineering Education at Princeton. His Massive Open Online Courses on networking reached over 250,000 students since 2012 and the textbook received the 2013 Terman Award from American Society of Engineering Education. He was named a Guggenheim Fellow in 2014. 
\end{biography}	
\balance
\newpage

\appendices
\section{Extension of users' optimization problem}
\label{sec:extension}
\revise{
	Our optimization problem in Section~\ref{sec:optProblem} assumes a constant speed for WiFi and prefers WiFi to 3G as long as WiFi is available. However, we can extend it to choose the WiFi speed for each application subject to a time- and location-varying maximum WiFi bandwidth (e.g., home WiFi is generally more stable than WiFi at a coffee shop). In this case, users can choose to use either 3G or WiFi instead of automatically using WiFi when it is available. Thus, users must account for the speed, cost, delay, and traffic amount for 3G and WiFi and choose the interface that yields the highest utility.
	The user's modified optimization problem is as follows:
	\begin{align}
	\max_{c_i^j(k,\gamma,\delta)}\;&\sum_{i = 1}^n\Bigg[\sum_{j\in J}\Bigg(\sum_{k\geq i}\Bigg(\sum_{\gamma\in\Gamma}\Bigg(\sum_{\delta\in\Delta}\Big(U_j\big(0, k - i, \delta, s_j(i)\big) +\nonumber \\
	&U_j\big(p, k - i, \gamma, s_j(i)\big)\Big)c_i^j(k,\gamma,\delta)\Bigg)\Bigg)\Bigg)\Bigg] \nonumber \\
	{\rm s.t.}\;p&\sum_{i = 1}^n\Bigg[\sum_{j\in J_v}\left(\sum_{k\geq i}\sum_{\gamma\in\Gamma} c_j^i(k,\gamma,0)s_j(i)\right) + \nonumber \\
	&\sum_{j\in J_t}\left(\sum_{k\geq i}\sum_{\gamma\in\Gamma} c_j^i(k,\gamma,0)\gamma s_j(i)\right)\Bigg]\leq B \nonumber \\
	&\sum_{i\leq l}\sum_{j\in J}\sum_{\gamma\in\Gamma} c_j^i(l,\gamma,0)\gamma\leq \beta\,\forall\,l \nonumber\\
	&\sum_{i\leq l}\sum_{j\in J}\sum_{\delta\in\Delta} c_j^i(l,0,\delta)\delta\leq \alpha(l)\,\forall\,l \label{eq:WiFiCapaConstraint}\\
	&\sum_{k\geq i} \sum_{\gamma\in\Gamma} \sum_{\delta\in\Delta} c_j^i(k,\gamma,\delta) = 1\,\forall\,j\in J;\;i = 1,2,\ldots,n \nonumber \\
	&c_j^i(k,\gamma,\delta) = 0,\forall\gamma>0,\delta>0 \label{eq:oneNetSelect} \\ 
	&c_j^i(k,\gamma,\delta)\in\left\{0,1\right\}. \nonumber
	\end{align}
	We change the indicator variables $c_i^j(k,\gamma)$ to $c_i^j(k,\gamma,\delta)$ so that WiFi speed $\delta$ can be chosen.
	The WiFi speed $\delta$ is chosen from a finite subset of possibilities $\Delta$. Here, $\Delta$ and $\Gamma$ include $0$ so that we can represent the case when the WiFi and 3G are not used. We add a capacity constraint on the WiFi bandwidth in (\ref{eq:WiFiCapaConstraint}). $\alpha(l)$ is the predicted WiFi bandwidth in period $l$. We limit the rate allocation so that only one network is chosen in (\ref{eq:oneNetSelect}). 
	We expect that the WiFi bandwidth prediction algorithm can be designed by modifying the Algorithm~\ref{alg:wifi} so that $v_k(l)$ represents the product of WiFi probability and bandwidth at time $k$ and location $l$, and the expected available bandwidth at each time is updated using empirical data.	
}

\revise{	
	This formulation can be modified into a second optimization problem to be solved in each period to choose the final 3G and WiFi bandwidths for applications scheduled to that period. For all application types $j$ originated in period $i$ where $\sum_{\gamma\in\Gamma}\sum_{\delta\in\Delta}c_i^j(k,\gamma,\delta)>0$ for the current period $k$, we solve following problem:}
  
\revise{	
	\begin{align}
	\max_{c_i^j(k,\gamma,\delta)}\;&\sum_{\gamma\in\Gamma}\Bigg(\sum_{\delta\in\Delta}\Big(U_j\big(0, k - i, \delta, s_j(i)\big) +\nonumber \\
	&U_j\big(p, k - i, \gamma, s_j(i)\big)\Big)c_i^j(k,\gamma,\delta)\Bigg) \nonumber \\
	{\rm s.t.}\;p&\sum_{\gamma\in\Gamma} c_j^i(k,\gamma,0)\gamma\leq \beta\, \nonumber\\
	&\sum_{\delta\in\Delta} c_j^i(k,0,\delta)\delta\leq \alpha(k)\, \nonumber \\
	&\sum_{\gamma\in\Gamma} \sum_{\delta\in\Delta} c_j^i(k,\gamma,\delta) = 1\, \nonumber \\
	&c_j^i(k,\gamma,\delta) = 0,\forall\gamma>0,\delta>0 \nonumber \\ 
	&c_j^i(k,\gamma,\delta)\in\left\{0,1\right\}. \nonumber
	\end{align}	  
	If the bandwidth constraints are violated, we can simply scale down the bandwidths assigned to different applications so as to satisfy the constrains as we discussed in footnote~\ref{footnote_constraint}. Instead, we may delay some applications to later periods, if we use the original formulation.   
}    
\section{Top 15 applications for WiFi and cellular networks}
\label{sec:top_applications}
Tables~\ref{tab:top_wifi} and ~\ref{tab:top_cellular} show the 15 most-used applications on WiFi and cellular networks, respectively.
\begin{table*}[ht]
	\renewcommand{\arraystretch}{1.1}
	\caption{Top 15 applications in WiFi network.} \label{tab:top_wifi}
	\vspace{-0.1in}
	\centering
	\begin{tabular}{|c||c||c|c|c|c|}
		\hline Index & Package name & Upload(\%) & Download(\%) & Total(\%) & Type \tabularnewline \hline \hline 
		
		1 & Streaming Media	&	70.20 	&	66.38 	&	68.14 	&	Video	\tabularnewline \hline
		2 & android.process.media	&	3.28 	&	3.26 	&	3.27 	&	Video	\tabularnewline \hline
		3 & com.google.android.music:main	&	3.22 	&	2.77 	&	2.98 	&	Unclassified	\tabularnewline \hline
		4 & com.google.android.music:ui	&	3.20 	&	2.75 	&	2.96 	&	Unclassified	\tabularnewline \hline
		5 & com.android.email	&	2.46 	&	2.59 	&	2.53 	&	Email	\tabularnewline \hline
		6 & com.emogoth.android.phone.mimi	&	2.67 	&	2.29 	&	2.46 	&	Social networking	\tabularnewline \hline
		7 & com.marvel.capinstaller	&	1.77 	&	1.52 	&	1.63 	&	Downloads	\tabularnewline \hline
		8 & com.android.browser	&	1.23 	&	1.69 	&	1.48 	&	Browsing	\tabularnewline \hline
		9 & com.clearchannel.iheartradio.controller	&	1.48 	&	1.27 	&	1.37 	&	Unclassified	\tabularnewline \hline
		10 & com.facebook.katana:providers	&	1.07 	&	1.30 	&	1.19 	&	Social networking	\tabularnewline \hline
		11 & com.facebook.katana	&	0.93 	&	1.31 	&	1.13 	&	Social networking	\tabularnewline \hline
		12 & com.motorola.process.system	&	0.05 	&	2.04 	&	1.12 	&	Unclassified	\tabularnewline \hline
		13 & com.rhythmnewmedia.android.e	&	0.81 	&	1.10 	&	0.97 	&	Unclassified	\tabularnewline \hline
		14 & com.ninegag.android.app	&	0.69 	&	0.65 	&	0.67 	&	Unclassified	\tabularnewline \hline
		& Others	&	6.93 	&	9.10 	&	8.10 	&-		\tabularnewline \hline
		
	\end{tabular}
\end{table*}
\begin{table*}[ht]
	\renewcommand{\arraystretch}{1.1}
	\caption{Top 15 applications in cellular network.} \label{tab:top_cellular}
	\vspace{-0.1in}
	\centering
	\begin{tabular}{|c||c||c|c|c|c|}
		\hline Index & Package name & Upload(\%) & Download(\%) & Total(\%) & Type \tabularnewline \hline \hline 
		
		1 & Streaming Media	&	5.70 	&	42.64 	&	33.77 	&	Video	\tabularnewline \hline
		2 & com.android.email	&	5.64 	&	5.59 	&	5.60 	&	Email	\tabularnewline \hline
		3 & android.process.media	&	4.03 	&	5.96 	&	5.50 	&	Video	\tabularnewline \hline
		4 & com.facebook.katana	&	7.47 	&	3.86 	&	4.73 	&	Social networking	\tabularnewline \hline
		5 & com.android.browser	&	6.78 	&	3.33 	&	4.16 	&	Browsing	\tabularnewline \hline
		6 & com.google.android.music:main	&	0.04 	&	5.27 	&	4.01 	&	Unclassified	\tabularnewline \hline
		7 & com.facebook.katana:providers	&	6.30 	&	2.51 	&	3.42 	&	Social networking	\tabularnewline \hline
		8 & com.rhythmnewmedia.android.e	&	5.26 	&	1.99 	&	2.78 	&	Unclassified	\tabularnewline \hline
		9 & com.pandora.android	&	4.69 	&	1.49 	&	2.26 	&	Unclassified	\tabularnewline \hline
		10 & com.noinnion.android.greader.reader	&	4.42 	&	1.40 	&	2.12 	&	Unclassified	\tabularnewline \hline
		11 & com.motorola.blur.service.main	&	2.89 	&	1.00 	&	1.45 	&	Unclassified	\tabularnewline \hline
		12 & com.motorola.contacts	&	2.89 	&	0.94 	&	1.41 	&	Unclassified	\tabularnewline \hline
		13 & com.alphonso.pulse	&	2.17 	&	1.07 	&	1.33 	&	Social networking	\tabularnewline \hline
		14  & com.motorola.process.system	&	0.35 	&	1.64 	&	1.33 	&	Unclassified	\tabularnewline \hline
		15 & com.clearchannel.iheartradio.controller	&	1.07 	&	1.25 	&	1.21 	&	Unclassified	\tabularnewline \hline
		& Others	&	40.28 	&	20.08 	&	24.93 	&-		\tabularnewline \hline
		
	\end{tabular}
\end{table*}


\end{document}